\documentclass[fleqn,usenatbib]{mnras}

\usepackage{newtxtext,newtxmath}
\usepackage{graphicx, caption}	
\usepackage{placeins}
\usepackage{amsmath}	
\usepackage{wasysym}
\usepackage{bm}          
\usepackage{color}
\usepackage{xspace}
\usepackage{mathtools}
\usepackage{esdiff}
\usepackage{ulem}
\usepackage{wrapfig}
\usepackage{lscape}
\usepackage{rotating}
\usepackage{xfrac}
\usepackage{relsize}
\usepackage{mathtools}
\usepackage{subcaption}
\usepackage[dvipsnames]{xcolor}
\usepackage{academicons}
\usepackage[export]{adjustbox}

\newcommand{\orcid}[1]{\href{https://orcid.org/#1}{\includegraphics[width=8pt]{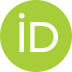}}}

\newcommand{\Msun}{\ensuremath{\,\mathrm{M}_\odot}\xspace}

\usepackage[T1]{fontenc}
\usepackage{ae,aecompl}

\definecolor{darkraspberry}{rgb}{0.53, 0.15, 0.34}
\definecolor{dodgerblue}{rgb}{0.12, 0.56, 1.0}
\definecolor{dogwoodrose}{rgb}{0.84, 0.09, 0.41}
\definecolor{cadmiumgreen}{rgb}{0.0, 0.42, 0.24}

\title[Models of massive contact binaries] {Detailed evolutionary models of massive contact binaries
\newline
I. Model grids and synthetic populations for the Magellanic Clouds}

\author[Menon et al.]{Athira Menon$^{\orcid{0000-0002-4234-4181}}$\,$^{1,2}$\thanks{E-mail: amenon@uni-bonn.de}, Norbert Langer$^{1,3}$, Selma E. de Mink$^{\orcid{0000-0001-9336-2825}}$\,$^{4,2,5}$, Stephen Justham$^{\orcid{0000-0001-7969-1569}}$\,$^{2,6,7,4}$, 
\newauthor
Koushik Sen$^{1,3}$, Dorottya Sz\'{e}csi$^{\orcid{0000-0001-6473-7085}}$\,$^{8,9}$,  Alex de Koter$^{\orcid{0000-0002-1198-3167}}$\,$^{2,12}$, Michael Abdul-Masih$^{\orcid{0000-0001-6566-7568}}$\,$^{10}$,
\newauthor
Hugues Sana$^{\orcid{0000-0001-6656-4130}}$\,$^{12}$, Laurent Mahy$^{\orcid{0000-0003-0688-7987}}$\,$^{11,12}$, Pablo Marchant$^{\orcid{0000-0002-0338-8181}}$\,$^{12}$\\
$^{1}$ Argelander-Institut für Astronomie, Universität Bonn, Auf dem Hügel 71, 53121 Bonn, Germany\\
$^{2}$ Anton Pannekoek Institute of Astronomy and GRAPPA, University of Amsterdam, Science Park 904, 1098 XH Amsterdam, Netherlands\\
$^{3}$ Max-Planck-Institut für Radioastronomie, Auf dem H\"ugel 69, 53121 Bonn, Germany\\
$^{4}$ Max Planck Institute for Astrophysics, Karl-Schwarzschild-Strasse 1, 85748 Garching, Germany\\
$^{5}$ Center for Astrophysics, Harvard-Smithsonian, 60 Garden Street, Cambridge, MA 02138, USA\\
$^{6}$ School of Astronomy and Space Science, University of the Chinese Academy of Sciences, Beijing 100012, China\\
$^{7}$ National Astronomical Observatories, Chinese Academy of Sciences, Beijing 100012, China\\
$^{8}$ Institute of Astronomy, Faculty of Physics, Astronomy and Informatics, Nicolaus Copernicus University, Grudzi\k{a}dzka 5, 87-100 Torun, Poland \\
$^{9}$ Physikalisches Institut, Universit\"{a}t zu K\"{o}ln, Z\"{u}picher-Strasse 77, D-50937 Cologne, Germany \\
$^{10}$  European Southern Observatory, Alonso de Cordova 3107, Vitacura, Casilla 19001, Santiago de Chile, Chile \\
$^{11}$Royal Observatory of Belgium, Avenue circulaire 3, B-1180 Brussel, Belgium\\
$^{12}$ Institute of Astrophysics, KU Leuven, Celestijnenlaan 200D, 3001 Leuven, Belgium.}

\date{Accepted XXX. Received YYY; in original form ZZZ}

\pubyear{2020}

\begin{document}
\label{firstpage}
\pagerange{\pageref{firstpage}--\pageref{lastpage}}
\maketitle

\begin{abstract}
The majority of close massive binary stars  with initial periods of a few days experience a contact phase, in which both stars overflow their Roche lobes simultaneously. We perform the first dedicated study of the evolution of massive contact binaries and provide a comprehensive prediction of their observed properties. We compute 2790 detailed binary models for the Large and Small Magellanic Clouds each, assuming mass transfer to be conservative. The initial parameter space for both grids span total masses from 20 to 80\Msun,  orbital periods of 0.6 to 2\,days and mass ratios of 0.6 to 1.0. We find that models that remain in contact over nuclear timescales evolve towards equal masses, echoing the mass ratios of their observed counterparts. Ultimately, the fate of our nuclear-timescale models is  to merge on the main sequence.  Our predicted period-mass ratio distributions of O-type contact binaries are similar for both galaxies, and we expect 10 such systems together in both Magellanic Clouds. While we can largely reproduce the observed  distribution, we over-estimate the population of equal-mass contact binaries.  This situation is somewhat remedied if we also account for binaries that are nearly in contact. Our theoretical distributions work particularly well for contact binaries with  periods $<2$\,days and total masses $\lessapprox45\Msun$. We expect stellar winds, non-conservative mass transfer and envelope inflation to have played a role in the formation of the more massive and longer-period contact binaries.

\end{abstract}

\begin{keywords}
   stars: massive -- stars: early-type -- binaries:close -- binaries: general -- stars: evolution -- stars: fundamental parameters
\end{keywords}



\section{Introduction}
\label{introduction}

A significant fraction of all stars in the Universe live in binary systems with companions that they are gravitationally bound to. Surveys of the Milky Way find that the fraction of low-mass stars found in such binary systems is 30--40\%  (e.g., \citealp{duquennoy1991,eggleton2008, raghavan2010,duchene2013}), while for massive stars this fraction goes up to 50--100\% (e.g., \citealp{vanbeveren1998, chini2012, sana2012, sana2014, dunstall2015, kobulnicky2014, aldoretta2015, moe2017}).

The largest homogeneous sample of O-type stars ($\textrm{M}\geq15\Msun$)  we have so far is from the young star-forming region, 30~Doradus (or Tarantula nebula), nestled in the Large Magellanic Cloud. The VLT-FLAMES Tarantula survey (VFTS, \citealp{evans2011}) obtained data for nearly 800 massive stars in 30~Dor, 360 of which are O-type stars \citep{sana2013} and 116 of which are binaries. A follow up study called the Tarantula Massive Binary Monitoring (TMBM) project calculated orbital solutions for nearly 82 of these systems \citep{almeida2017, mahy2020a, mahy2020b}, thus providing us with a rich, unbiased database to study the lives of O stars in binary systems. 

A large fraction of early-type massive binaries in both, the Galactic and the VFTS samples, have orbital periods of a few days. These systems are expected to evolve into a contact configuration  \citep{wilson2001}, wherein both stars overflow their Roche lobes simultaneously and become bound by a common equipotential surface that extends between the inner and outer Lagrangian points of either star, lending them their unique peanut-shaped structure.

Investigating the evolution of massive contact binaries is essential  to estimate the fraction of the binary population that will merge on the main sequence or evolve further until they  explode or collapse to black holes. Those that do merge before either star forms a compact object may give rise to fast-rotating single stars which are considered to be progenitors of long duration $\gamma-$ray bursts \citep{yoon2005, Woosley:2006, yoon2006, Meynet:2007, Dessart:2008, vanMarle:2008, Szecsi:2017long, AguileraDena:2018},  B[e] and sgB[e] stars (e.g., \citealp{podsiadlowski2006, vanbeveren2013, justham2014, wu2020}), blue supergiant progenitors of Type II supernovae such as SN~1987A \citep{podsiadlowski1992, menon2017, urushibata2018, menon2019}, progenitors of  superluminous supernovae \citep{justham2014, Sukhbold:2016, aguilera-dena2020}, pulsational pair instability supernovae \citep{Langer:1991, Heger:2003, langer2007, Woosley:2007, Chatzopoulos:2012, gomez2019} and magnetic stars \citep{schneider2016,schneider2019}.

Low-mass contact binaries,  called W UMa binaries,  have been observed quite frequently-- data is available for hundreds of these systems, and they have orbital periods between 0.3 and 1\,day, mass ratios between 0.1 and 0.8, and total masses of approximately $1\Msun$ \citep{szymanski2001, selam2004, rucinksi2007b}. In comparison, fewer massive contact binaries are known from observations. Surveys of O\& B-type stars have so far reported nearly 40 such systems in the Large Magellanic Cloud (LMC), 2 such systems in the Small Magellanic Cloud (SMC) and 17 in the Milky Way. They have total masses from 17 to 85\Msun, orbital periods from 0.45 to 6.6\,days and mass ratios between 0.3 and 1. The Magellanic Cloud contact binaries are much less wider than those in the Milky Way, and are also more confined in the period-mass ratio space than the more scattered Milky Way  systems.

Observationally, contact binaries are identified by their distinct light curve shapes. Unfortunately in most systems, the signature of contact is not always clear-cut since their degree of contact is uncertain, and could imply that they may only be nearing contact and not be in contact at present. In some cases the same system has a different status assigned by different works that studied them. We will look at these uncertainties in more detail later in the paper. Prominent among the handful of systems which are definitively in deep (over) contact, is VFTS\,352 in 30~Dor, one of the most massive contact binaries known. The timescale over which the orbital period of this over-contact system varies is $\dot{P}/P\approx0.6$\,Myr, indicating an ongoing nuclear-timescale mass transfer \citep{almeida2015}.  Similar time derivatives have also been inferred for several other massive contact binaries as well (e.g., \citealp{qian2006a,qian2007, martins2017}). The majority of these deep contact binaries have equal-mass components, including VFTS\,352. 

Despite the data available for massive contact binaries, there are at present no known detailed   models that thoroughly investigate their evolution or explain their observed stellar and orbital parameter distribution. In contrast, there have been several evolutionary models for low-mass contact binaries which explain their observed orbital period and mass ratio distribution and, their spread along the main sequence (e.g., \citealp{webbink1976,yakut2005,stepien2006, gazeas2008,stepien2012,zhang2020}). An important aspect of these models is the heat transfer in their common convective envelope, which leads to large-scale circulations of mass and thermal energy in the envelope and helps explain the temperature and luminosity differences between the component stars of the binary  \citep{lucy1968, flannery1976, webbink1977, stepien2009}.

Contrary to the low-mass contact binaries, massive contact binaries share a common envelope that is largely radiative and the majority of them are located close to ZAMS. The physics of their common radiative envelope is not well understood, especially the heat transfer that may occur between their components, and is hence not included while calculating the evolution of massive contact binaries.  

There have been some exploratory calculations for massive contact binaries, the earliest among them published in the series of papers by Sybesma \citep{sybesma1985,sybesma1986a,sybesma1986b}. The next set of works such as those of \citet{nelson2001}, \citet{wellstein2001} and \citet{demink2007} explored the parameter space in which binaries can enter contact during Case A mass transfer and the impact of semi-convection and mass-transfer efficiency on the evolution of their models. \citet{demink2007} computed a large grid of binary models at the SMC metallicity and identified which main-sequence binaries will undergo contact and which will avoid contact. They also evolved some models that were in shallow contact. \citet{vanbeveren1998, mennekens2017} also investigated the evolution of contact binaries, but only of thermal-timescale contact binaries. They stopped the evolution of those models that were expected to enter contact over a nuclear-timescale, as they expected these systems to merge eventually. Of particular relevance to our work are the works of \citet{marchant2016,marchant2017}, who computed grids of binary models for a range of metallicities (Z$_\odot$/4...Z$_\odot$/50) and initial binary parameters. We will discuss their work shortly.

An important open question concerning the evolution of close binaries, is the efficiency of internal mixing due to rotational instabilities.  Rotationally-induced mixing is predominantly due to the circulation of meridional currents in the outer radiative layers of a star \citep{eddington1926,eddington1929}  and is expected to become stronger with increasing rotational velocity, increasing initial mass, and decreasing metallicity (e.g.,\citealp{heger2000a,heger2000b, maeder2009, brott2011a, brott2011b}.) Mixing in the radiative outer layers causes the transport of nuclear-burning products such as helium and nitrogen from the convective core to the surface of the star. In its most extreme form, rotational mixing can cause the star to undergo chemically homogeneous evolution (CHE), wherein mixing is so efficient that the stars lose their core-envelope stratification soon after ZAMS, and remain hot and compact throughout their evolution. This channel of evolution is considered to be particularly important in the formation of rapidly-rotating metal-poor single stars \citep{maeder1987, yoon2005,yoon2006,brott2011a,szesci2015}.

In the case of binaries, CHE has been suggested to occur in very short-period systems on the main sequence, in which the stars become deformed by tides \citep{deMink2009, song2013, song2016, hastings2020}. In such systems, this type of evolution can prevent the stars from merging and they evolve as massive (and compact) helium stars which end their lives as heavy black holes \citep{deMink2009, song2016, deMink2016, mandel2016, marchant2016, dubuisson2020}. CHE has also been considered for near-contact binaries \citep{deMink2009, song2016, riley2020} and over-contact binaries \citep{marchant2016}.  In particular, the models computed by \citet{marchant2016, marchant2017}  were found to undergo CHE during long contact phases, as their metallicity decreased (Z$<\textrm{Z}_\odot$/2), primary masses increased (M$_{1}\geq40\Msun$) and for initial mass ratios closer to unity.

This paper is intended as the first in a series that will have the ultimate goal of (1) improving our understanding of the physics of massive contact binaries, (2) gaining insight in their evolutionary pathways and observable properties and (3) providing reliable predictions for their final fate as stellar mergers or possibly massive double helium stars and gravitational-wave progenitors.   

In this first paper we focus on very close binaries and follow their evolution through contact on the main sequence.  We perform stellar evolution calculations for a large grid of short-period binaries varying their initial masses, orbital periods and mass ratios, with an initial composition that is typical for young stars in the LMC and SMC. We study the evolution of massive binaries from ZAMS and through the contact phase and, in particular, the changes in their orbital properties prior and during the contact phase. We make predictions for the distributions of their orbital parameters and provide a first comparison with observed massive contact systems.  We show where our predictions agree well with observations, but also discuss tensions that exist that can provide us with further insight. 

Our paper is organized as follows: Section~\ref{methods}
describes our model assumptions, the initial parameter space of the grids and our method for computing orbital parameter distributions. There are two results sections: Section~\ref{results1} discusses how massive contact systems typically form and evolve in our grid and how the choice of initial parameters affect their evolution.  Section~\ref{results2} compares the distributions calculated from our models with the observed distributions, including the number of massive contact binaries we expect in the Magellanic Clouds. In Section~\ref{discussion} we present a discussion of our results and point out avenues where we can improve them, and in Section~\ref{conclusions}, we summarize our findings and indicate directions for future work.

\section{Methods}\label{methods}
To compute a large grid of binary-star models suitable for comparison with observed systems in the LMC and SMC, we use version 10398 of the 1D stellar evolution code \texttt{MESA} \citep{paxton2011, paxton2013, paxton2015, paxton2018, paxton2019} \footnote{The necessary files to reproduce our results with this version of \texttt{MESA} are available at \url{https://github.com/athira11/massive_contact_binaries.git}}. In the remainder of this section we discuss our physical assumptions (Section~\ref{physics}), the initialization and termination of our models (Section~\ref{initialization})
and our assumptions for modelling a population of contact binaries (Section~\ref{distributions}).

\subsection{Stellar physics assumptions}
\label{physics}
\subsubsection{Mixing, mass loss, microphysics}
Various mixing processes are implemented as time-dependent diffusive processes in \texttt{MESA}, where the user has the freedom to vary their efficiencies. Convection is modelled using the standard mixing-length theory \citep{bohm1958,cox1968} prescription with a mixing length parameter of $\alpha_\textrm{MLT}=1.5$, following \citet{pols1998}. We adopted the Ledoux criterion for convective stability \citep{ledoux1958}. We allow for mixing beyond the edge of convectively unstable regions by assuming a step overshooting parameter  $\alpha_\textrm{ov}=0.335$ as calibrated by \citet{brott2011a}. Semiconvection is included with an efficiency parameter of $\alpha_\textrm{sc}=1.0$ as in \citet{langer1983}. This value of this parameter can significantly affect the evolution of the accreting binary components \citep{braun1995}. \citet{schootemeijer2019} find a good
agreement between stellar models and the distribution of massive stars in the
Hertzsprung-Russel diagram of the SMC for $\alpha_\textrm{sc}=1.0$. Thermohaline mixing is included by adopting a value of 1.0 for the dimensionless parameter $\alpha_\textrm{th}$.
The sources of rotationally-induced mixing and angular momentum transport considered in \texttt{MESA} include the Eddington-Sweet circulation, secular and dynamic shear instabilities, and the Goldreich-Schubert-Fricke instability \citep{heger2000a,heger2005}. The efficiency factor for the total rotationally-induced mixing coefficient is set to f$_\textrm{c}=1/30$ as determined theoretically by \citet{chaboyer1992}. The factor controlling the inhibition of rotational mixing against gradients in mean molecular weight is set to f$_\mu=0.1$ as in \citet{yoon2006}. Angular momentum transport due to magnetic fields is implemented as in \citet{heger2005} and \citet{petrovic2005}, which is motivated by observations (e.g.,\citealp{suijs2008}).

 Stellar winds are implemented depending on the surface helium mass fraction ($Y_\textrm{s}$) as in \citet{yoon2006}; if $Y_\textrm{s}<0.4$, i.e., for hydrogen-rich stars, mass-loss rates are computed as in \citet{vink2001}, while for hydrogen-poor Wolf-Rayet stars (with $Y_\textrm{s}>0.7$) the \citet{hamann1995} prescription is used but reduced by a factor of 10. For both rates, we use a metallicity-dependent wind whose strength is proportional to $Z^{0.86}$. For stars with $0.4<Y_\textrm{s}<0.7$, the mass-loss rates are interpolated between the above two prescriptions.
 
Nucleosynthesis reactions are calculated using the \texttt{MESA approx21.net} network containing 21 isotopes from $^1$H until $^{56}$Ni and encompasses the necessary reactions for the hydrogen CNO burning cycle, thereby allowing us to follow  the evolutionary phase of interest. Radiative opacities are calculated using CO-enhanced opacity tables from the OPAL project \citep{iglesias1996}.

\subsubsection{Binary star physics assumptions}
We use the binary module of \texttt{MESA} as described in \citet{paxton2015} and \citet{marchant2016} to compute the evolution of interacting binary models in our grid. Tidal synchronization is implemented according to the \citet{hut1981} and \citet{hurley2002} prescriptions. Mass transfer due to Roche lobe overflow is implicitly calculated using the Ritter scheme \citep{ritter1988} and is switched to the contact scheme when both stars overfill their Roche lobes, as will be discussed in Section~\ref{contact}. As we are studying close binaries in this work, we assume that both stars in the system are synchronized to their initial orbital period at the beginning of the evolution but we allow for differential rotation. 

Mass transfer is treated conservatively except due to mass loss through winds. We apply the standard mass transfer physics of \texttt{MESA} \citep{paxton2015} which computes the mass transfer rate through an implicit scheme, including angular momentum accretion and tidal spin-orbit coupling. The kinetic energy of the accretion stream is neglected \citep{ulrich1976}
and the entropy of the accreted material is assumed to be the same as that of the surface of the mass gainer.

\subsubsection{Treatment of the contact phase}
\label{contact}

Mass transfer during contact, i.e., when both stars in a binary system simultaneously overflow their Roche-lobe volumes, is implemented in \texttt{MESA} as described in \citet{marchant2016} and \citet{marchant_thesis}. We summarize the main features of the contact scheme here.

During contact we assume that the surfaces of both stars lie on a common equipotential surface. The amount of mass transferred from one star to the other is adjusted such that the following relationship holds: 

\begin{align}
\ \frac{R_{2}(\Phi)-R_\textrm{RL,2}}{R_\textrm{RL,2}} = F(q,\frac{R_{1}(\Phi)-R_\textrm{RL,1}}{R_\textrm{RL,1}})
\label{RLOF}.
\end{align}
where $\Phi$ is the equipotential surface shared by both stars, R$_\textrm{j}$ and R$_\textrm{RL,j}$ are the volume-equivalent stellar and Roche lobe radius respectively of either star (j=1,2) and $q=M_2/M_1$ where $M_2$ is the current less massive star in the binary. The function $F(q,x)$, where $x=\frac{R_{1}(\Phi)-R_\textrm{RL,1}}{R_\textrm{RL,1}}$, is solved by numerically integrating the equipotential volume of each star through the inner Lagrangian point L1 for different mass ratios. $F(q,x)$ is thereby approximated as $F(q,x)=q^{0.52}x$ for $x>0$ such that $F(q,x)=0$ if $x=0$ and is equal to $x$ if $q=1$, i.e., both stars have equal radii when they attain equal masses.

The volume equivalent radius corresponding to the outer Lagrangian point L2 (R$_\textrm{L2,2}$) depends on the Roche lobe radius of the less massive star (R$_\textrm{RL,2}$) and the mass ratio $q$ at the time considered, and is calculated as:

\begin{align}
\ \frac{R_\textrm{L2,2}-R_\textrm{RL,2}}{R_\textrm{RL,2}} = 0.299\, \textrm{tan}^{-1} \left (1.84\,q^{0.397} \right).
\label{L2OF}
\end{align}

Energy transport in the common envelope is not accounted for during the contact phase. Note that in this subsection (and this subsection only) we follow the notation used by \citet{marchant2016} where subscript 1 refers to the star that is the more massive one at a given time during the evolution of the binary model.  This may be the primary (the initially more massive star) or the secondary (the initially less massive star) if the mass ratio has been reversed as a result of mass exchange. Throughout the rest of this paper we will use the common convention where 1 (2) refers to the  \textit{initially} more (less) massive star or the primary (secondary) star.

\subsection{Initialization and termination of models}

\subsubsection{Initial parameters}
\label{initial_parameters}

We compute a total of 2790 binary models for each galaxy grid and begin the evolution of each model with both stars on the ZAMS (we describe this initialization in Section~\ref{initialization}). The initial parameter range of our grids and the grid spacing ($\Delta$ values) are:
\begin{itemize}
    \item \textbf{Initial mass ratio}:
    q$_\textrm{i}\equiv\textrm{M}_\textrm{2,i}/\textrm{M}_\textrm{1,i}$ = 0.6, 0.7, ...,1.0 with $\Delta$q$_\textrm{i}=0.1$ and we add an extra grid for q$_\textrm{i}=0.95$.
    \item \textbf{Initial total mass}: M$_\textrm{T,i}\equiv \textrm{M}_\textrm{1,i} + \textrm{M}_\textrm{2,i}$ = 20, 22, ...,80\Msun with $\Delta$M$_\textrm{T,i}=2$\Msun
    \item  \textbf{Initial period}: P$_\textrm{i}=0.6, 0.7, ...,2.0$\,days with $\Delta$P$_\textrm{i}=0.1$\,days
    \end{itemize}

We restrict the lower limit of the initial mass ratio to q$_\textrm{i}=0.6$ as convergence errors become more common for lower initial mass ratios and the upper limit of the total mass to M$_\textrm{T,i}=80\Msun$ as the observed contact binaries have total masses less than this value (except for one).  Our choice to space our grid in equal steps of the initial total mass, M$_{\rm T,i}$, instead of the more usual choice of the initial primary mass M$_{\rm 1,i}$ is a natural one when considering contact binaries that experience near conservative mass transfer.  The total mass stays approximately constant during the early evolution of these models while the masses of the individual components can change substantially as the system experiences one or more phases of mass transfer (cf.,  Fig~\ref{evol1}). We note that our choice of sampling the grid evenly in total mass and mass ratio leads to effects at the edges of our grid towards the lowest and highest initial primary masses (M$_{\textrm{1,i}}<12.5$\Msun and M$_{\textrm{1,i}} > 40$\Msun, respectively) which are only possible for a limited combination of choices of the mass ratio. However, these `edge effects' are found to be very small and do not affect our main conclusions. 

The range in initial orbital period is motivated to cover the systems of interest for the formation of long-lived contact systems. Systems with shorter periods are already in contact at zero age and merge immediately. Although systems with wider periods may contribute, especially for the most massive systems that start with almost equal masses, we expect that their contributions do not significantly affect our main conclusions at the metallicities of our grids. \citet{sen2021} also report a similar finding for binary models with initial orbital periods above 2 days.

Our choice of initial parameter space is different from that of \citet{marchant2016}, who were mainly interested in the parameter space in which binaries can evolve chemically homogeneously and form double black hole systems. They hence evolved very low-metallicity massive binary models with initial primary masses between 25--502\Msun and initial mass ratios equal to 1 (except for their most metal-poor grid). In comparison, we explore much smaller masses with initial mass ratios from 0.6--1 in our grids, as we intend to cover the parameter range of the spectroscopic contact binaries identified in the Magellanic Clouds. In addition, our method of initializing our binary models is also different from \citet{marchant2016}, as we explain in the next section.

\subsubsection{Initialization of models}
\label{initialization}
Before initiating the binary evolution of models in our grid, we first build a separate library of starting models. For each binary model we compute single star ZAMS models of masses M$_\textrm{1,i}$ and M$_\textrm{2,i}$. We define the ZAMS model as the point of evolution at which a stellar model has contracted from the pre-main sequence (pre-MS) branch to its smallest radius before expanding on the MS.

This initial setup differs from that of \citet{marchant2016} where the binary evolution is initiated by allowing \texttt{MESA} to find the appropriate pre-MS models for the required masses and initial composition. In their setup, the two stellar models are puffed up at the beginning of the evolution before shrinking and expanding again during their MS evolution.  As a consequence, the stars may interact prior to reaching the ZAMS point by transferring mass and thereby changing the orbital period before their MS expansion actually begins. 
Some of the models in this setup of \citet{marchant2016} even merge during the transition from the pre-MS to MS branch. 

The `initial' masses and orbital period of the binary models presented in this paper refer to the masses of the primary and secondary stars and the orbital period of the binary system at the beginning of the ZAMS. By using carefully constructed starting models for the primary and secondary stars, our initial setup allows us to provide more direct predictions of the evolutionary trends of close binaries. We also find that more binary models avoid merging initially on the ZAMS and their overall lifetime on the MS is longer compared to the models from the setup of \citet{marchant2016} since fewer of them come in contact at zero age.

For the initial composition we follow \citet{brott2011a} and choose a chemical mixture that is representative of stars in the SMC and LMC, with Z$_\textrm{SMC}=0.0021$ and Z$_\textrm{LMC}=0.0047$ which are $\approx \textrm{Z}_{\odot}/2$ and $\approx \textrm{Z}_{\odot}/5$ respectively, where Z$_{\odot}=0.014$ \citep{asplund2005}. The individual abundances of 7 elements: H, He, C, N, O, Mg, Si and Fe are included as reported in Table~1 of \citet{brott2011a} and the abundances of all other elements are scaled-down to the solar abundances of \citet{asplund2005}.

For each binary model in a grid corresponding to a given metallicity, we need three main inputs to begin their evolution: their initial period and the ZAMS models corresponding to the individual masses, M$_\textrm{1,i}$ and M$_\textrm{2,i}$. At the start of the simulation we assume the binary is tidally synchronized, hence within a short timescale (less than 0.01\,Myr), the rotational periods of the stars become equal to the orbital period of the system.

\vspace{-0.5cm}
\subsubsection{Outcomes}

Since we are only modelling the MS binary phase of each model, the evolution is terminated when either star leaves the MS or when  the system undergoes L2 overflow.
The latter is expected to result in mass loss through the outer Lagrangian point and a corresponding loss of angular momentum  from the binary, leading to a common envelope phase and the eventual merger of the binary system. Our models experience one of these three final outcomes:

\begin{enumerate}
    \item \textbf{Overflows L2 initially}: The stars are initially so close that the  binary system experiences mass loss through the L2 Lagrangian point while the stars have barely evolved (the secondary has depleted less than 3\% of its central hydrogen abundance). 
    \item \textbf{Overflows L2 on MS}: If the binary system experiences overflow through the L2 point post ZAMS but while still on the MS, they are denoted as `Late L2 overflow' systems. These systems merge while still burning hydrogen in their cores. 
    
    \item \textbf{Survive the MS}: If one or both stars evolves past the MS, we call these `MS survive' binaries. 

\end{enumerate}

Of the 2790 models in the LMC grid, nearly 13\% faced convergence errors i.e., the binary evolution was unable to initiate correctly or terminated sooner than their proper end (one of the three outcomes above). The contribution from these non-convergent models to our synthetic populations is negligible since they fall in the parameter space where they will either experience L2 overflow at the start of the simulation or will undergo contact only on a thermal timescale ($<0.01$\,Myr). Of the models that do converge, nearly 16\% of them experience `initial L2 overflow' and 63\% of them `overflow L2 on MS'; thus nearly 80\% of our LMC models merge on the MS. These fractions are similar for the SMC grid of models as well.

\subsection{Modelling the population of contact binaries distributions}
\label{distributions}

Each evolutionary model is characterized by a unique combination of initial orbital period P$_\textrm{i}$, initial total mass M$_\textrm{T,i}$
and initial mass ratio q$_\textrm{i}$ (and hence a choice of M$_\textrm{1,i}$) sampled from the parameter space as described in Section~\ref{initial_parameters}. The birth distributions are given by:

\begin{align}
\frac{\textrm{dN}}{\textrm{dM}_\textrm{1,i}} \sim \, \textrm{M}_\textrm{1,i}^{\alpha}, \, \frac{\textrm{dN}}{\textrm{dq}_{\textrm{i}}} \sim\, {\textrm{q}_{\textrm{i}}}^{\kappa}, \,
\frac{\textrm{dN}}{\textrm{dP}_{\textrm{i}}}& \sim\, {\textrm{P}_{\textrm{i}}}^{\gamma},   
\end{align}

For the purpose of a first order calculation, we assume the initial mass function (IMF) has an exponent of $\alpha = -2.35$ according to the Salpeter IMF \citep{salpeter1955, kroupa2002}. We also assume an {\"Opik} law in period, i.e., flat in log\,P (which is $\gamma=-1$ for P), and a flat distribution in q, i.e., $\kappa = 0$. The birth weight of a given  initial choice of M$_\textrm{1,i}$, P$_\textrm{i}$ and q$_\textrm{i}$ is then calculated as:

\begin{subequations}
\label{eq:weight}
\begin{align}
w_{\textrm{M}_{1,\textrm{i}}} &= \textrm{C}_\textrm{M$_1$}\,\left[\textrm{M}_\textrm{1,i}^{-1.35}\right]_{\textrm{M}_\textrm{1,i}-\textrm{dM1,u}}^{\textrm{M}_\textrm{1,i}+\textrm{dM1,l}} \\
w_{\textrm{P}_{\textrm{i}}} &= \textrm{C}_\textrm{P}\,\left[\textrm{log$_{10}$\,P}_\textrm{i}\right]_{\textrm{P}_\textrm{i}-\textrm{dP}}^{\textrm{P}_\textrm{i}+\textrm{dP}} \\
w_{\textrm{q}_{\textrm{i}}} &=
\textrm{C}_\textrm{q}\,\left[\textrm{q}_\textrm{i}\right]_{\textrm{q}_\textrm{i}-\textrm{dq}}^{\textrm{q}_\textrm{i}+\textrm{dq}},
\end{align}
\end{subequations}
where $\textrm{C$_\textrm{M$_1$}$}, \textrm{C$_\textrm{P}$}$, and $\textrm{C$_\textrm{q}$}$ are normalization factors multiplied with their respective weights to ensure that the sums of $w_{\textrm{M}_{1,\textrm{i}}}$, $w_{\textrm{P}_{\textrm{i}}}$ and $w_{\textrm{q}_{\textrm{i}}}$ are each equal to 1. 

The mass, period, and mass ratio intervals  in Eqs.~\ref{eq:weight} (dM$_1$, dP, dq) define the grid cell dimensions for each considered binary evolution model in the initial parameter space. 

 For the initial period, $\textrm{dP}=0.05$\,days, which is half the grid spacing. For the mass ratio, we choose $\textrm{dq}=0.05$ for all choices of q$_\textrm{i}$ except for the highest mass ratios where our models are spaced more densely.  We adapt the cell boundaries such that they lie exactly in between our choices of 
 q$_\textrm{i}$. In particular, we choose dq for our q$_\textrm{i}=1.0$ grid of models such that they represent the systems between q$_\textrm{i}=0.975$ and q$_\textrm{i}=1$. While our grid is evenly spaced in total mass (see Section~\ref{initial_parameters}), the spacing in initial primary mass is not even as M$_\textrm{1,i}= \textrm{M}_\textrm{T,i}/(1+\textrm{q}_\textrm{j})$ for any combination of i and j. Therefore, we define the upper and lower limits for M$_\textrm{1,i}$ as: 
\begin{align}
\label{eq:dm}
\mathrm{dM_\textrm{1,u}}= \frac 
{(\textrm{M}_\textrm{1,i+1}
-\textrm{M}_\textrm{1,i})}{2}\Msun,\,\,
\mathrm{dM_\textrm{1,l}}=\frac{(\textrm{M}_\textrm{1}
-\textrm{M}_\textrm{1,i-1})}{2}\Msun.
\end{align}
where the cell size does depend on the mass ratio q$_\textrm{j}$.  However, with this definition, the parameter space is covered without gaps or overlaps. For the primary masses at the edge of the grids, we use: at M$_\textrm{T,i}=20\Msun$, $\mathrm{dM_\textrm{1,l}}=0$ and at M$_\textrm{T,i}=80\Msun$, $\mathrm{dM_\textrm{1,u}}=0$ while for all other primary masses the limits are calculated as in Eq.~\ref{eq:dm}.

Finally, the statistical weight $w_\textrm{s}$ of a model `s' with its particular combination of the above initial parameters, which is the convolution of the birth weight and the dimension of the cell it represents, is:
\begin{align}
\label{eq:initialw}
w_\textrm{s} & = w_{\textrm{M}_{1},\textrm{i}} \times w_\textrm{q,i} \times w_\textrm{P,i}.
\end{align}

\subsubsection{Probability distributions of contact binaries according to their initial parameters}
\label{initial_hist}

One of our goals is to study the distribution of the fraction of time systems spend in contact compared to their overall MS lifetime, weighted by their birth weights. We define this fractional contact time per system in two ways: 

\begin{align}
\boldsymbol F_\textrm{1,s}={w_\textrm{s}} \frac{\tau_\textrm{contact,s}}{\tau_\textrm{MS-binary,s}},   \,\, \boldsymbol F_\textrm{2,s}= {w_\textrm{s}} \frac{\tau_\textrm{contact,s}}{\tau_\textrm{MS-single,s}},  \label{eq:F2} 
\end{align}
where $\tau_\textrm{contact,s}$ is the net time spent in contact by the system and $w_\textrm{s}$ is the birth weight of the system as calculated in Eq~\ref{eq:weight}. $\tau_\textrm{MS-binary,s}$ is the lifetime of the binary model before it merges on the MS or before either of its components leaves the MS. In practice since many of these close binaries merge before they exhaust their central hydrogen, we also define $\tau_\textrm{MS-single,s}$, which is the MS lifetime of a star had it not been part of a binary system that is close enough to interact. We take $\tau_\textrm{MS-single,s}$ for each binary system to be the lifetime of a single star of 1/2 the total binary mass, M$_\textrm{T,i}$. The values of $\tau_\textrm{MS-single,s}$ for each M$_\textrm{T,i}$/2 in our grids, were supplied from the grid of single star models computed by Schootemeijer, A. via private communication. The SMC set of these models are published in \citet{schootemeijer2019}.

We then compute the overall fraction of contact systems among the MS binaries in our population, $\boldsymbol f_\textrm{contact/MS}$. Using this number we can predict the number of massive contact binaries in a given population of MS binaries. It is calculated as follows: 

\begin{align}
\ \boldsymbol f_\textrm{contact/MS}=\frac{\sum _{\textrm{s=1}}^{\textrm{n}}{w_\textrm{s} \tau_\textrm{contact,s}}} {\sum _{\textrm{s=1}}^{\textrm{n}}{w_\textrm{s} \tau_\textrm{MS-binary,s}}} \label{eq:f_contact_MS},
\end{align}
where s is an index that loops through all 2790 binary models in each grid. 

A detailed explanation for how we compute the distributions shown in this paper can be found in Sections~\ref{initial_dist} and ~\ref{contact_dist}.

\section{Results}\label{results1}

In the following section, we will describe the properties of our binary evolution models, which will be mainly
presented in the form of distribution functions. We start by discussing two of our models in detail, which are representative of the evolutionary paths followed by the majority of contact binary models in our study. 

\subsection{Example models}
\label{examples}
\begin{figure*}
    \centering
   \includegraphics[width=\linewidth, clip]{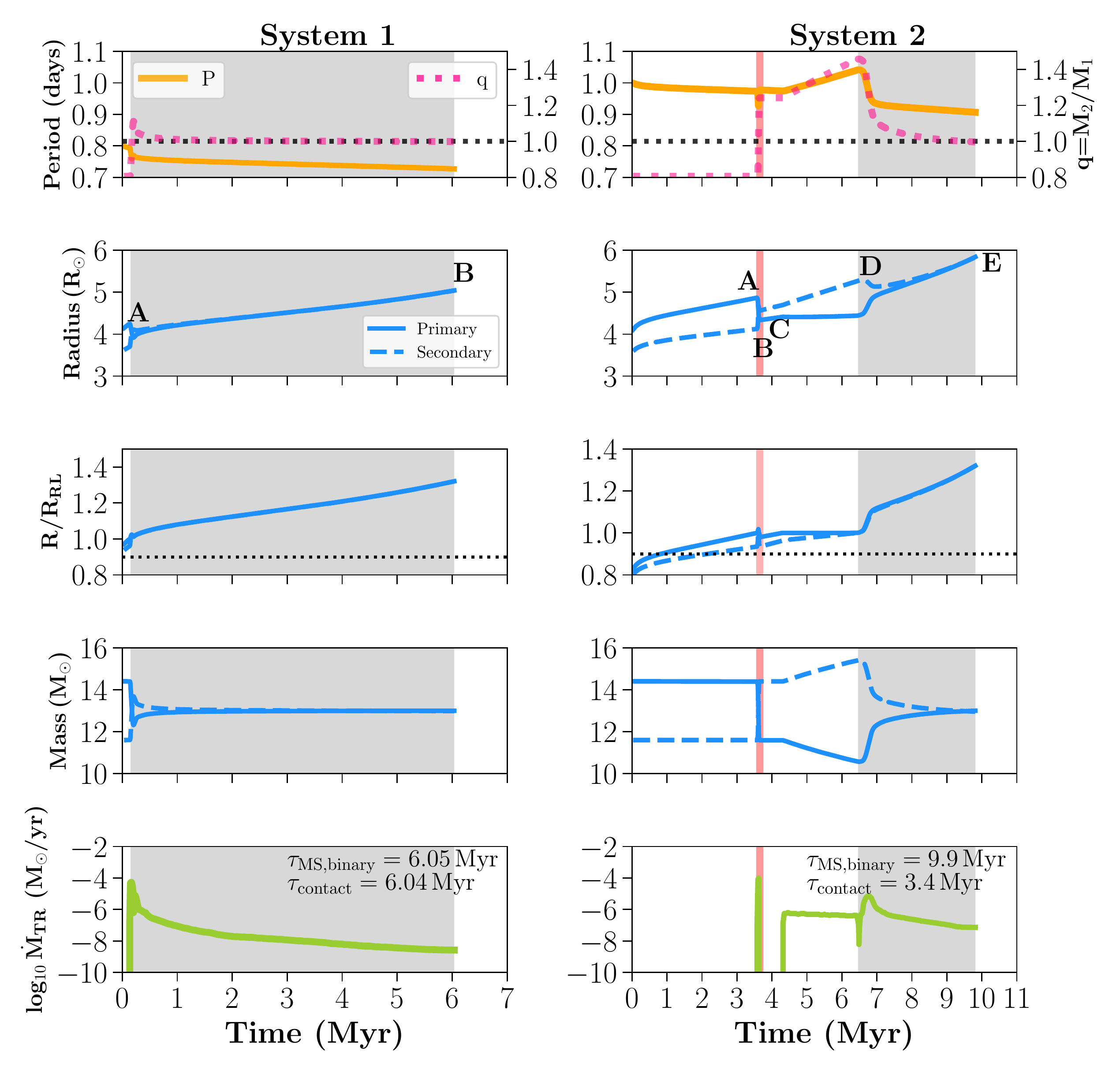} \caption{The evolution of the two example systems described in Section~\ref{examples}, with initial masses of 14.4 and 11.6\Msun for the primary (solid blue lines) and secondary (dashed blue lines) respectively. System 1 (left panels) begins with P$_\textrm{i}=0.8$\,days and System 2 (right panels) begins with P$_\textrm{i}=1.0$\,days. In the topmost panels, the left y-axis represents the period evolution and the right y-axis the mass ratio evolution. The black dotted horizontal line in the topmost panels is a reference for $\textrm{q}=1.0$ while in the third panel, the dotted line indicates when both stars have R/R$_\mathrm{L}>=0.9$. The letters in each panel correspond to the respective phases of each system, as described in Section~\ref{examples}. The true contact phases (where both stars have R/R$_\mathrm{L}>=1$) are shaded in grey (nuclear timescale) and red (thermal timescale).}
    \label{evol1}
\captionsetup{width=0.5\textwidth}
\end{figure*}

\begin{figure*}
\centering
\begin{subfigure}{0.48\textwidth}
  \centering
  \includegraphics[width=\linewidth,left]{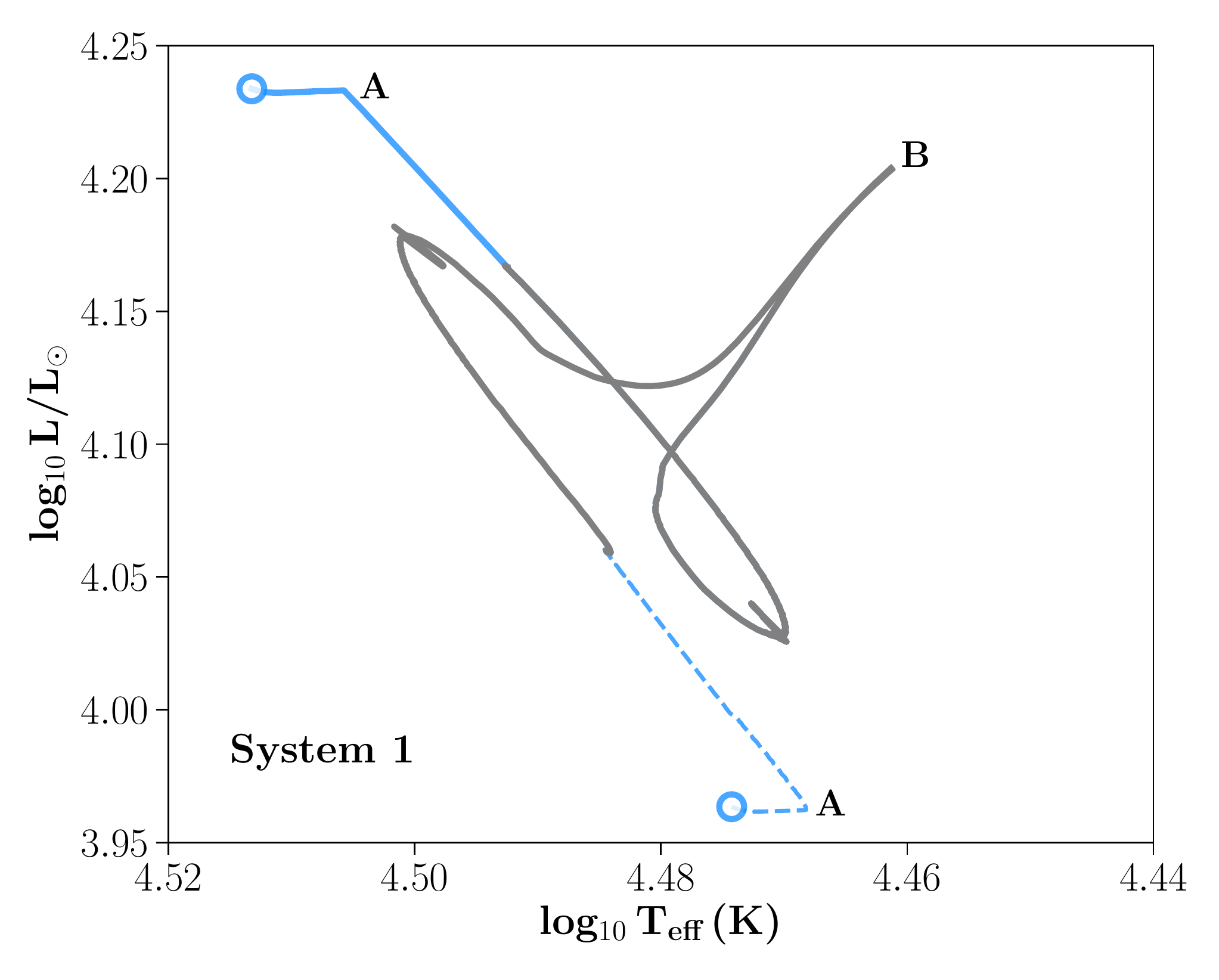}
\end{subfigure}%
\begin{subfigure}{0.48\textwidth}
  \centering
  \includegraphics[width=\linewidth]{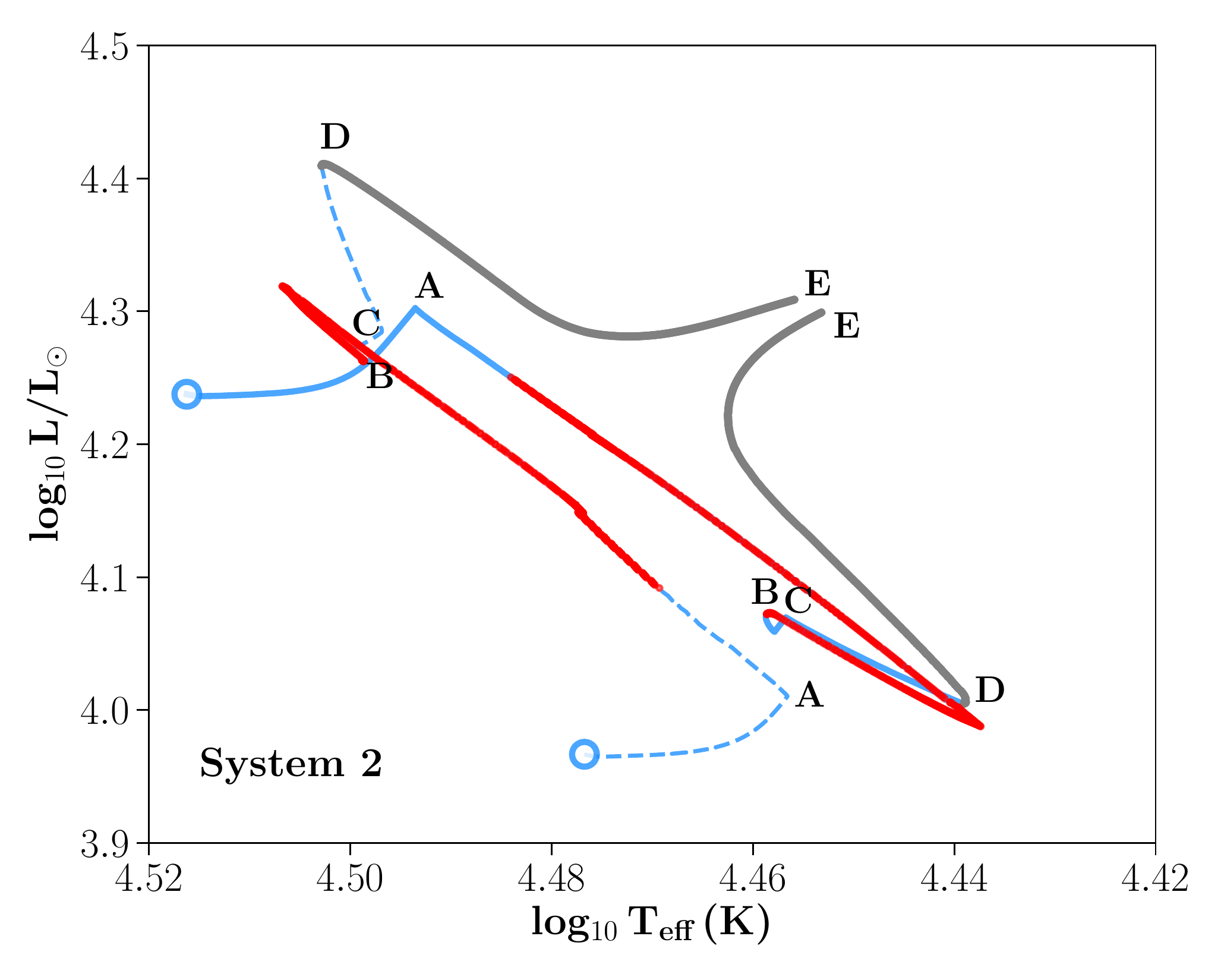}
\end{subfigure}
\caption{HR diagram of the two systems from Fig.~\ref{evol1}, including the phases of evolution marked by their corresponding letters (see text in Section~\ref{examples}). Solid line represents the primary star and dashed line represents the secondary star. The grey region denotes the nuclear timescale contact phase and the red region denotes the thermal timescale.}
\label{HRD_models}
\end{figure*}

\begin{figure*}
\centering
   \includegraphics[width=\linewidth,clip]{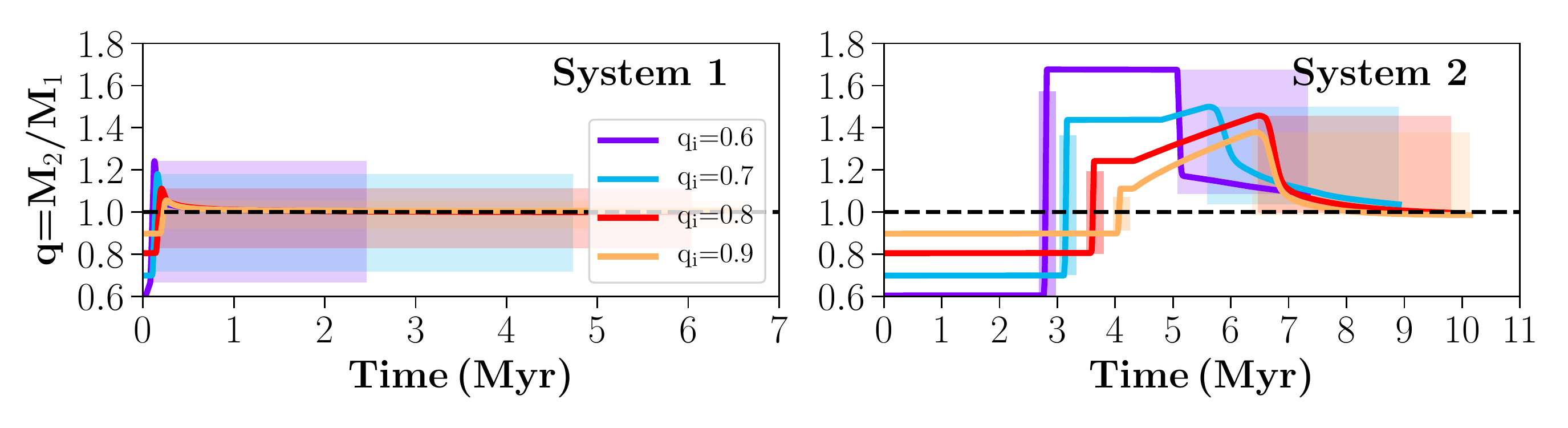}
    \caption{The impact of the initial mass ratio on the evolution of binary systems that come in contact.  We show the mass ratio q as a function of time for systems 1 and 2 from Fig.~\ref{evol1}, for a fixed total mass M$_\textrm{T,i}=26$\Msun and respective initial periods (0.8\,days and 1.0\,day) but with different initial mass ratios q$_\textrm{i}=0.6,0.7,0.8,0.9$. The colored shaded regions indicates the contact phase corresponding to each q$_\textrm{i}$. The dashed black line indicates $\textrm{q}=1$.}
    \label{evol2}
\end{figure*}

 \begin{figure}
\includegraphics[width=\linewidth, left,clip]{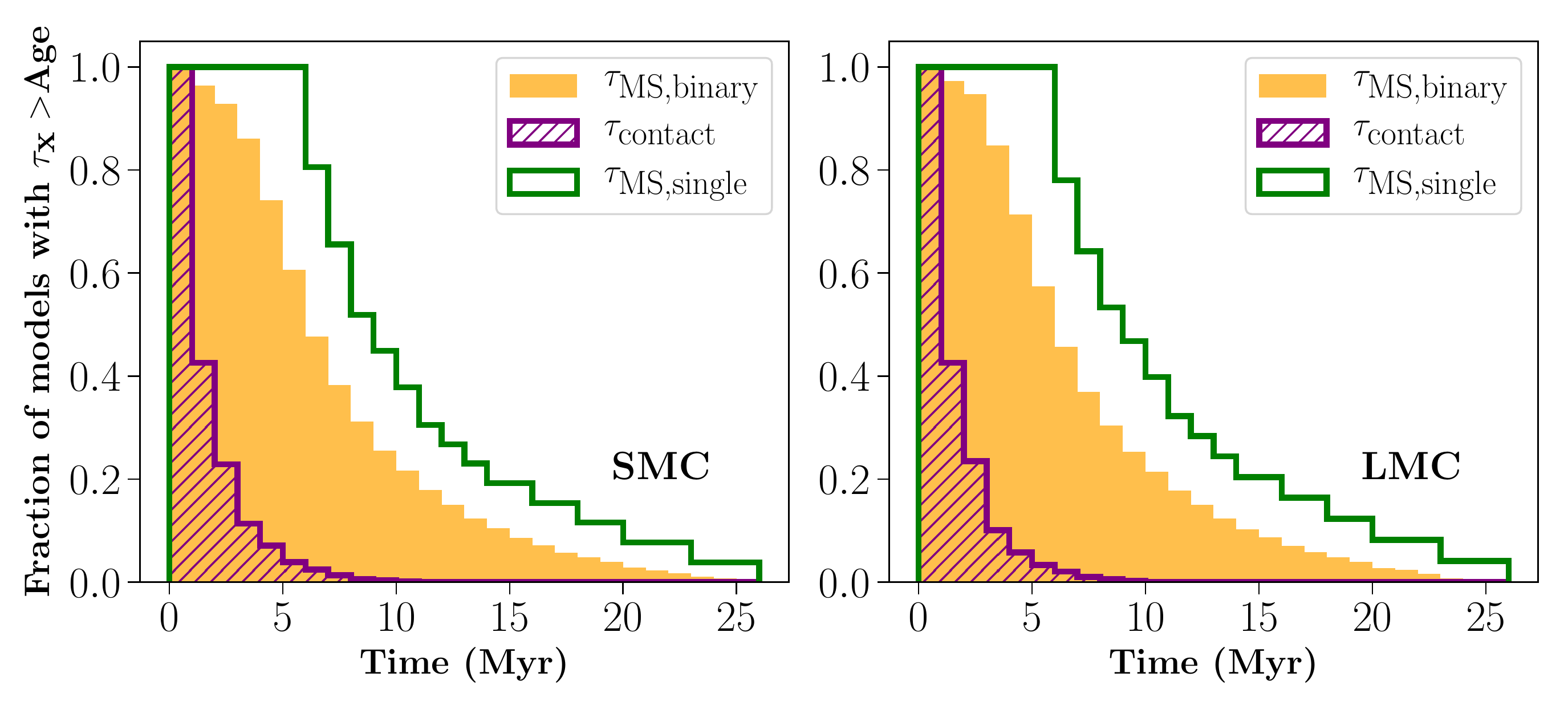}
\caption{Reverse cumulative distribution of three timescales of all models in the SMC grid (left plot) and LMC grid (right plot), of the contact phase
$\tau_\textrm{contact}$ (hatched purple histogram), $\tau_\textrm{MS,single}$ (orange filled histogram) and $\tau_\textrm{MS,binary}$ (green histogram). See Section~\ref{initial_hist} for their definitions. Only those models which have $\tau_\textrm{MS,binary}\geq 0.1$\,Myr are included in these histograms.}
\label{age_hist}
\end{figure}

\begin{figure}
\captionsetup{format = hang, width=\linewidth}
\includegraphics[width=\linewidth, left,clip]{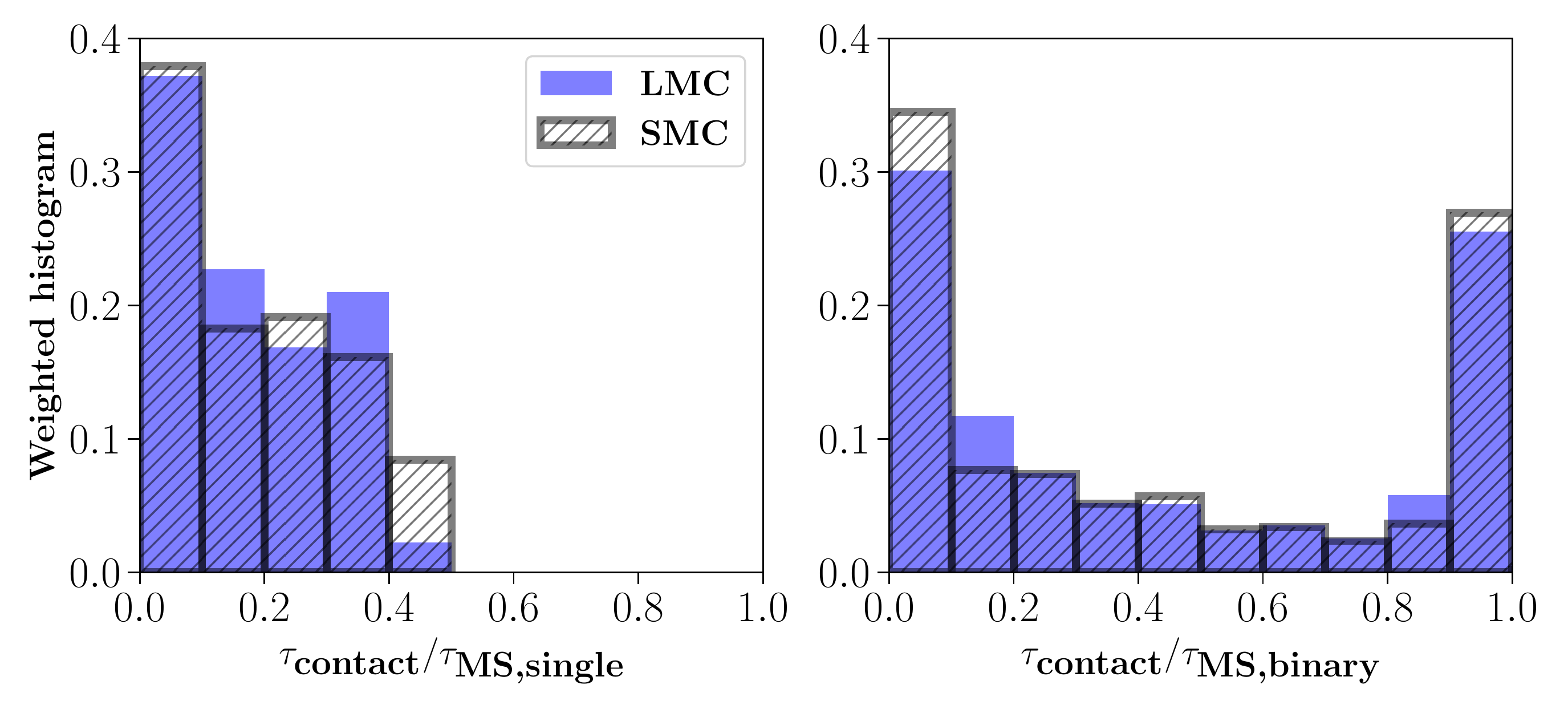}
\caption{Weighted histogram of fractional contact duration for all the models in the SMC (grey striped histogram) and LMC (purple filled histogram), as a fraction of  the equivalent single-star MS lifetime (left plot) and the binary model MS lifetime (right plot). Each model is weighted by its birth parameters. As in Fig.\,\ref{age_hist}, we do not include models with $\tau_\textrm{MS,binary}<0.1$\,Myr.
}
\label{frac_hist}
\end{figure}

The binary models in our grids go through at least two of three mass transfer episodes which occur in the following order: i) contact on a thermal timescale ii) semi-detached phase \textit{or} second contact on a thermal timescale iii) contact on a nuclear timescale. The two example models we use to illustrate the major evolutionary channels in our grids only differ in their initial periods, but otherwise have the same initial mass ratio of q$_\textrm{i}=0.8$ and total mass of M$_\textrm{T,i}=26$\Msun. `System 1' has an initial period of 0.8\,days and `System 2' has an initial period of 
 1.0\,day. As defined in Section~\ref{contact}, the primary is  the \textit{initially} more massive star and the secondary is the \textit{initially} less massive star. Please refer to Figs.~\ref{evol1} and \ref{HRD_models} (left panels for system 1 and right for system 2) in tandem with this section.
 
 The first episode of mass transfer in system 1 occurs soon after initializing the binary evolution. A fast Case  A mass transfer occurs from the primary to the secondary for the first $\approx0.01$\, Myrs. The orbital period dips slightly during this mass transfer phase and then rises again when the mass ratio reverses, as is expected during conservative Case A mass transfer (point A, left panel set of Fig.~\ref{evol1}). Stars with radiative envelopes typically shrink in radii as they lose mass and experience a drop in their surface luminosities and effective temperatures, while the reverse happens when they gain mass. These effects can be seen in the trajectory of the primary and secondary stars on the HR diagram in Fig.~\ref{HRD_models}. Thereafter, the more massive secondary transfers mass on a nuclear timescale and the system enters contact. The orbital period decreases as the system converges towards equal masses. Both stars expand on the MS and remain in contact as a slow mass transfer continues from the secondary until it overflows the L2 Lagrangian point at about 6\,Myr, after which the evolution is terminated (point B). Models that evolve like system 1 are in contact for most of their binary lifetimes before merging on the MS. Models that evolve like system 1 have the longest contact durations in both our metallicity grids and their contribution dominate the synthetic binary populations we will see in later sections.
  
Since system 2 begins initially wider than system 1, its first mass transfer episode occurs at about $3.6$\,Myr. The system enters contact and transfers mass on a thermal timescale (points A to B, right panel set of  Fig.~\ref{evol1}). The primary loses mass and shrinks in radius while the secondary gains mass and expands (right panel, Fig.~\ref{HRD_models}). The orbital period drops  and rises again in this fast mass-transfer episode, and both stars detach and break contact. Both stars continue to expand and at nearly 4.3\,Myr, a second mass transfer phase ensues from the now less-massive primary on a nuclear timescale, and the system enters a semi-detached configuration for nearly 3\,Myr  (points C to D). As the donor is the less massive star, the orbital separation widens during this mass transfer phase. The mass-transfer scheme employed by \texttt{MESA} during the semi-detached phase fixes the radius of the donor to its Roche lobe radius such that the donor radius remains constant during this phase. In the meantime, the secondary gains mass and continues to expand but does not overflow its Roche lobe until nearly 6.5\,Myr. At this point, the third mass transfer episode ensues with both stars overflowing their Roche lobes and the system thus enters contact.  

The mass transfer during this contact phase occurs from the now more massive secondary to the less massive primary. Initially the reverse mass transfer rate is high, reaching about $10^{-5} \Msun\,\rm{yr}^{-1}$ but it gradually decreases to a nuclear timescale and the mass ratio steadily approaches unity as the orbital period decreases. This type of evolution is reminiscent of the typical rapid and slow phases of Case A systems described in the literature, albeit with significant differences.  The initial rapid phase lasts longer than the thermal timescale that is typically seen in standard Case A systems and the mass ratio does not (re)reverse in this system after it achieves $\textrm{q}=1$. After nearly $3.4$\,Myr in this slow contact phase, the system experiences L2 overflow (point E) with a final mass ratio very close to unity. Some models in our grids that evolve like system 2 experience L2 overflow before attaining a mass ratio of 1. In such cases, the final mass ratio of the system prior to merger is close to, but slightly  different from 1.  

It is not only important to consider the contact phase where both stars overflow their Roche lobe volumes, but also phases where both stars are nearly in contact. As mentioned in Section~\ref{introduction}, observationally it can be difficult to distinguish  over-contact systems from those approaching contact. We assume that this  `\textit{near-contact phase}' in binaries, occurs when  both stars simultaneously have R/R$_\textrm{L}\geq0.9$, while not yet achieving contact (R/R$_\textrm{L}\geq1$ for both stars). Models that follow the evolution of system 2 contain an extended phase of being in near contact during the semi-detached phase (between points B and C) of their evolution. While the actual contact phase of system 2 lasts 3.4\,Myr, it spends an additional\,3.2 Myr in near contact. We hence introduce the definition of the \textit{`relaxed contact phase'}, which is the sum of the near and actual contact phases of a system, during which it may be classified as a ``true" contact system. In the case of system 2, the duration of this relaxed contact phase is 6.6\,Myrs.

The effect of including these relaxed-contact binaries in our synthetic stellar populations significantly impacts our predictions for the observed contact binary population, as we shall see in Section~\ref{results2}.

\subsection{Impact of the initial mass ratio on systems 1 and 2}
\label{initial_q}

In general, the initial mass ratio (q$_\textrm{i}$) of the models in our grids determine the duration of their $\tau_\textrm{MS,binary}$ and $\tau_\textrm{contact}$; the closer their q$_\textrm{i}$ is to 1, the longer they are in contact before merging. 

Fig.~\ref{evol2} shows the evolution of systems 1 and 2 but now with varying initial mass ratios.  Independent of q$_\textrm{i}$, models that evolve like system 1 enter contact almost immediately after ZAMS and attain a mass ratio of unity within about 1\,Myr, staying in contact until the end of their MS evolution (see also \citealp{marchant2016}). 

In contrast, the final mass ratio of models that evolve like system 2 does depend on their initial mass ratio. The further their initial mass ratios are further away from 1, the shorter the duration of their contact phase and the time they spend as equal-mass contact binaries.  While all the models asymptotically approach $\textrm{q}=1$, the majority of models with q$_\textrm{i}=0.6$ and 0.7  experience L2 overflow prior to attaining $\textrm{q}=1$.  With decreasing q$_\textrm{i}$ the duration of the semi-detached phase, which is responsible for the second increase in mass ratio of the models, becomes smaller and altogether disappears in the q$_\textrm{i}=0.6$ model. This is because the system widens considerably after the first thermal-timescale mass transfer episode and remains detached for about 2\,Myrs. The next mass transfer phase occurs from the massive secondary on a thermal timescale thereby reversing the mass ratio from $\textrm{q}=1.65$ to 1.2 and the system enters its second contact phase. Thereafter, the mass transfer rate slows down to the nuclear timescale of the expanding secondary and undergoes L2 overflow before attaining a mass ratio of unity.

In both types of evolution, the net contact duration decreases as the initial mass ratio decreases. This is because the expansion of the primary star drives the evolution of the system; the more massive the primary is initially, the sooner the system attains contact and more rapid its overall evolution. In the case of models that evolve like system 2, the dip in orbital period corresponding to the  (re)reversal of the mass ratio (during the second rapid mass transfer phase) becomes more pronounced as the initial mass ratio decreases, causing the system to merge sooner as well.

We summarize our findings thus far: a mass ratio of one is the equilibrium configuration that all our binary models proceed towards when they come in contact over a nuclear timescale. As their initial mass ratio increases, the more time the binary model spends in contact and the likelier it is to attain a final mass ratio of one before merging.

\subsection{A census of the contact binary models in our grids: their life-span and fates.}

\subsubsection{The duration of the contact phase in our models}

\begin{figure*}
\centering
  \includegraphics[width=0.8\textwidth]{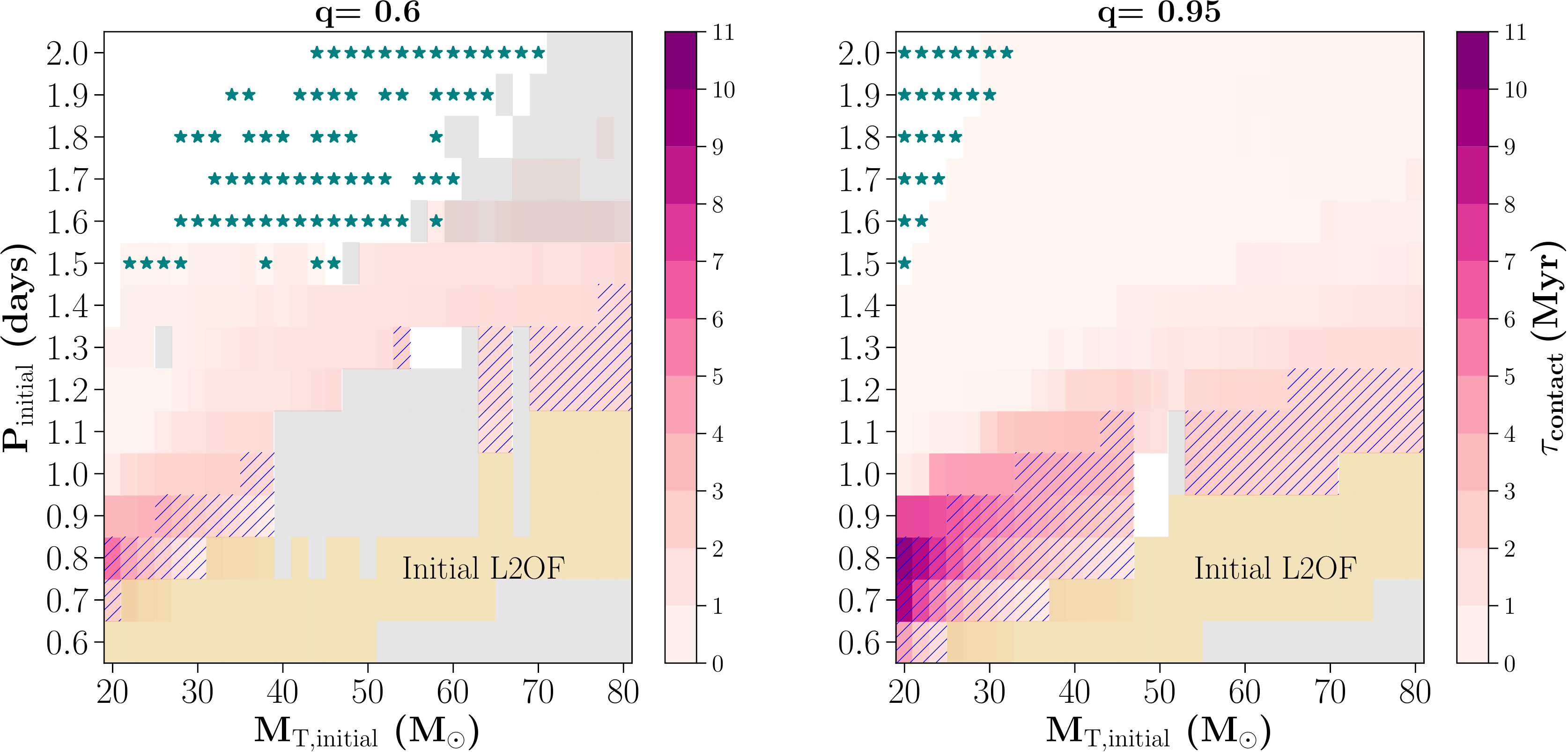}
\caption{Contact duration ($\tau_\mathrm{contact}$) of our LMC models as function of their
initial orbital period (P$_\textrm{initial}$) and initial total mass (M$_\textrm{T,initial}$)
for two initial mass ratios, q$_\mathrm{i}=0.6$ (left) and q$_\mathrm{i}=0.95$ (right). The background colors represent the contact duration of each system in the grid; the more purple a pixel is, the longer the contact phase of the corresponding system is. The parameter space colored in yellow undergo Initial L2 overflow while the one colored in grey experienced convergence errors. Grey pixels in the lower half of the diagram are also expected to experience Initial L2 overflow. Pixels with blue hatching represent models that evolve as system 1, while those are that are neither hatched nor shaded in grey or yellow, represent models that evolve like system 2.  Systems with white backgrounds spend less than 0.01\,Myr in contact.  Green star symbols represent `MS survive' models, i.e., the system avoids merging during core hydrogen burning. All other systems experience L2 overflow 
and merge during their MS evolution.}
\label{scatter_sub}
\end{figure*}

\begin{figure*}
\centering
\begin{subfigure}{0.48\textwidth}
  \centering
  \includegraphics[width=\linewidth,left]{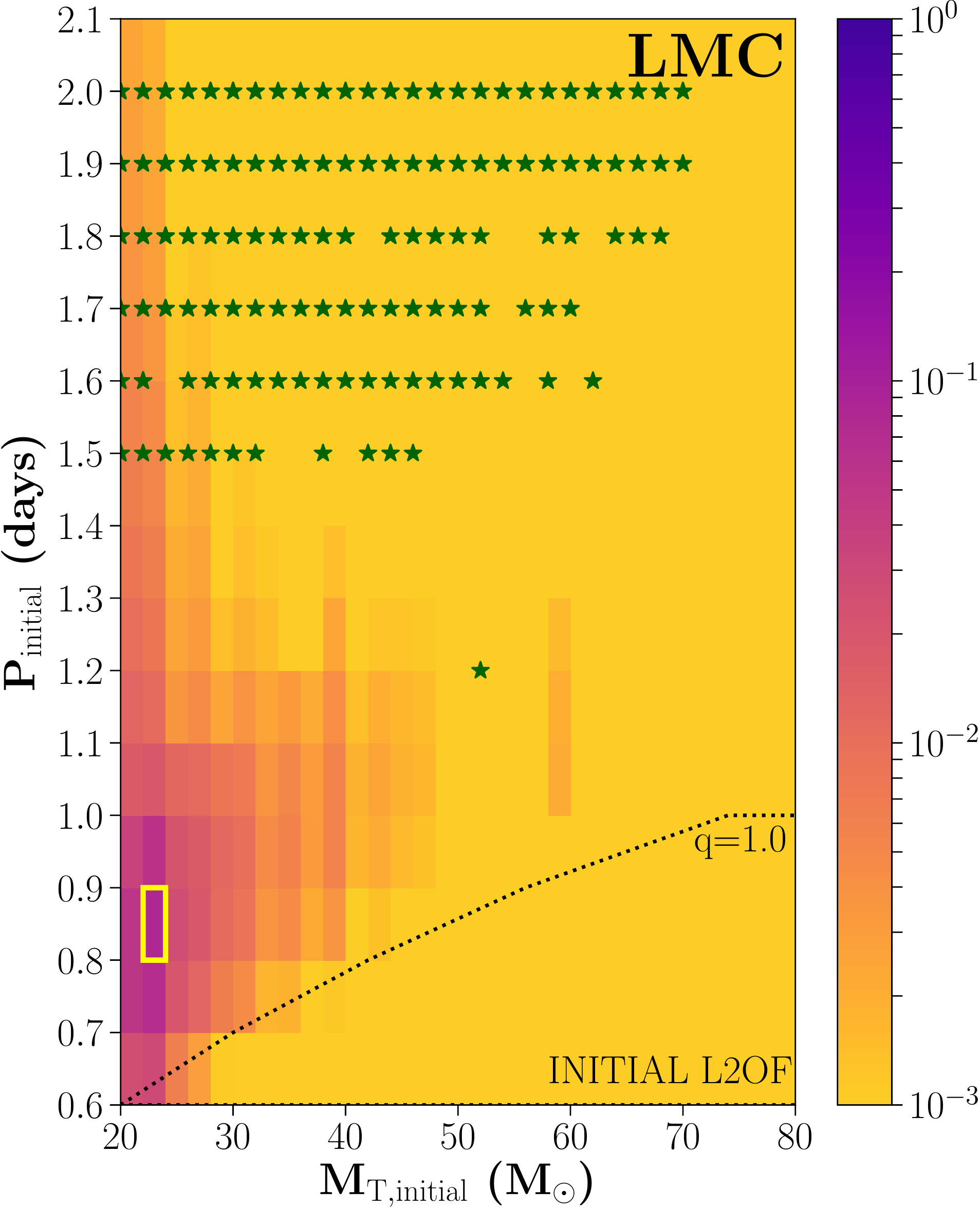}
\end{subfigure}%
\begin{subfigure}{0.48\textwidth}
  \centering
  \includegraphics[width=\linewidth]{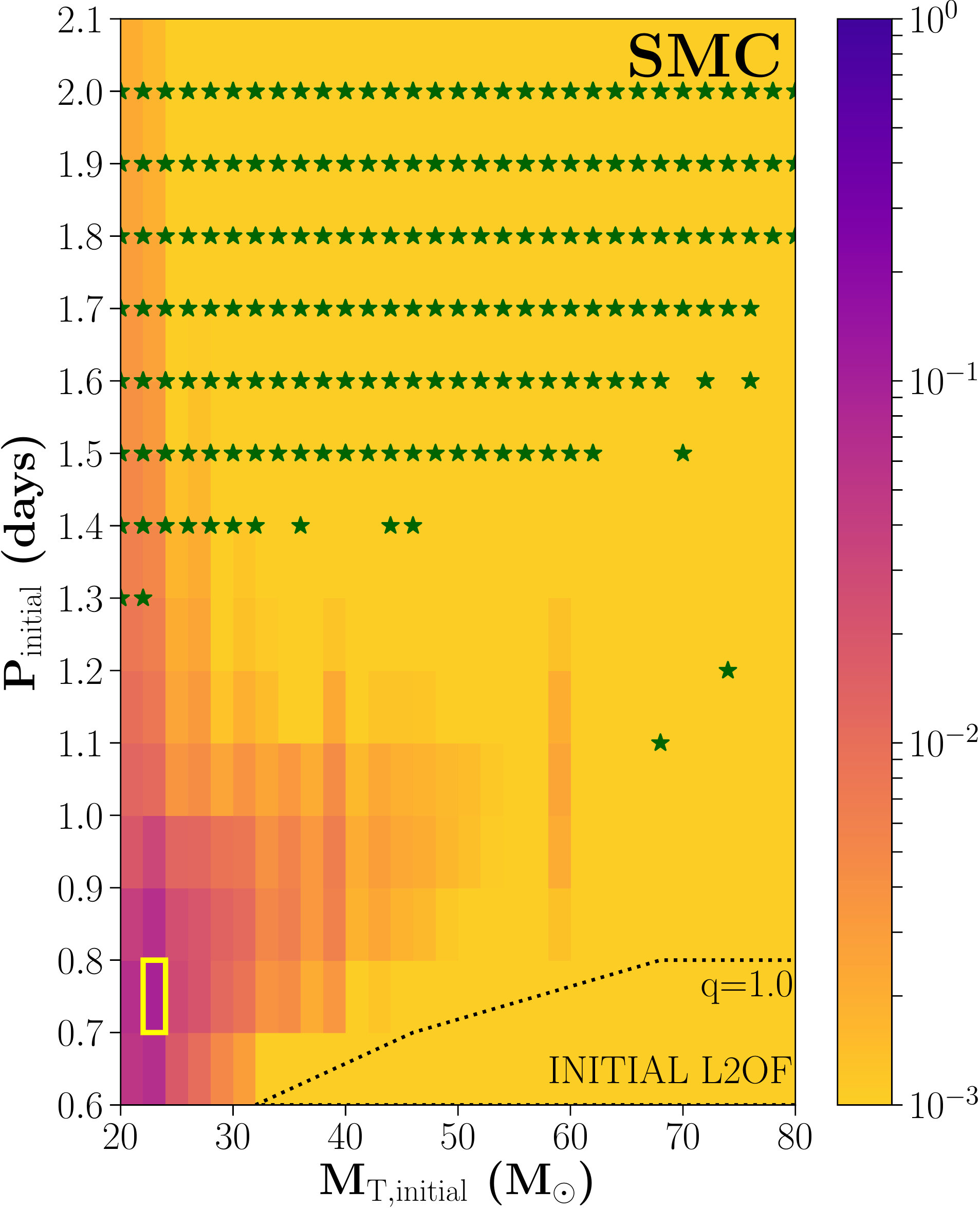}
\end{subfigure}
\caption{2D probability distribution showing the contributions of systems to a synthetic contact
binary population as function
of their initial period (P$_\textrm{initial}$) and initial total mass (M$_\textrm{T,initial}$), across all initial mass ratios. The background color of each pixel is the normalized probability of a binary with a particular initial orbital period and total mass contributing to the contact binary population (cf., Section~\ref{initial_dist}). The yellow rectangular box emphasizes the pixel which provides the largest contribution. Turquoise star symbols indicate pixels which contain binary models which avoid merging during the MS evolution. The area below the dotted black line indicates those systems that experience initial L2OF for an initial mass ratio of q$_\textrm{i}=1.0$.}
\label{P_MT_initial}
\end{figure*}

We briefly examine the duration of the contact and MS phases of the binary models in our grid. Figure~\ref{age_hist} shows the contact duration for the models as an unweighted cumulative histogram, along with their MS life time before merging ($\tau_\textrm{MS, binary}$) and their equivalent lifetime had they been single stars ($\tau_{\rm MS, single}$), which is the MS lifetime of a star with half the total mass of the binary system (cf., Section~\ref{initial_hist}). For the sake of clarity, we exclude systems that merge within 0.1\,Myr ($\tau_\textrm{MS,binary}\geq 0.1$\,Myr).

We see that about 65\% of our models experience contact for less than 1\,Myr. These systems are born in contact or very close to contact and merge soon after.  The net contact duration of a system also decreases when its total mass increases, owing to their shorter overall nuclear timescale, or with initial mass ratios further from 1 (cf., Section~\ref{initial_q}). The longest contact durations are $10.1\,$Myr and $10.7\,$Myr in the LMC and SMC grid respectively, which belong to the models with the lowest initial total mass and an initial mass ratio of one. 

Figure~\ref{frac_hist} shows the distribution of the fractional contact durations, normalized to their MS durations and weighted by their birth parameters. On average, the weighted contact lifetimes of our models is about 45\% of their MS binary lifetimes, for both the LMC and SMC (right plot in Fig.~\ref{frac_hist}). We also see that the weighted distribution of $\tau_\textrm{contact}/\tau_\textrm{MS,binary}$
has two peaks, at 0-10\% and at 90-100\%. While the first peak corresponds to the similar peak in the left plot of Fig.~\ref{frac_hist} (between 0 and 0.1), the second peak is produced by systems that remain in contact for most of their MS binary lifetimes similar to system 1 or merge shortly after being born in contact (but with $\tau_\textrm{MS,binary}>0.1$\,Myr). To discern the contributions of the latter type of models that merge soon after ZAMS, the normalisation to equivalent single star lifetimes, $\tau_{\rm MS, single}$, gives a better idea about the expected number of contact binaries in a population of mixed single stars and binaries (see Section~\ref{conclusions} below). From Fig.~\ref{frac_hist} we find that our models are in contact for at most 50\% of their single-star MS lifetime.  

We further note that the distributions in Figs.~\ref{age_hist} and \ref{frac_hist} for our SMC and LMC models are remarkably similar, despite a difference of a factor of 2.5 in their metallicity. This is an indication that metallicity is not a primary factor of importance, at least in the parameter range studied here. 

Figure~\ref{scatter_sub} shows two slices of our three dimensional LMC grid with the contact duration as a function of the initial period (y-axis) and initial total mass (x-axis) for the models with the most extreme mass ratios: q$_\textrm{i}=0.6$ and q$_\textrm{i}=0.95$. We do not include the q$_\textrm{i}=1$ models in this analysis as they are special and will be discussed later in this section. 

 Systems that experience L2 overflow at initialization (yellow squares in Fig.~\ref{scatter_sub}) are expected to merge quickly and not contribute to the observable population of contact binaries. The fraction of models with convergence issues (grey shaded squares) increases for total masses more than 40\Msun and periods shorter than 1.3\,days, where we typically also find L2 overflow at the start of the evolution. This fraction increases for our more extreme initial mass ratios, where we experience problems at wider periods for the higher total masses. 

 Models that evolve similar to our example system 1 (blue hatched squares) become less common as the total mass increases (for a fixed period) and as the period increases (for a fixed total mass). For the lowest total mass systems in our grid, models follow a system 1 like evolution for initial periods up to 0.8\,days for the least massive models in our grid, and initial periods between 1.1 and 1.5\,days for the most massive systems in our grid. These models spend at least 70\% of their MS binary lifetimes as contact systems and have the longest contact durations in our grid. They also attain equal masses before merging during core hydrogen burning. 

 Models that evolve in a similar way as our example system 2 are very common, and typically come from models with initial periods larger than 0.8\,days and become more common as q$_\textrm{i}$ decreases. These models evolve through a semi-detached phase and may merge while they still have unequal masses.

Models in which at least one star evolves past the MS (the `MS survive systems', marked as green stars in Fig.~\ref{scatter_sub}) have orbital periods larger than 1.5\,days and become increasingly common as the total masses and initial separation increases.  For the more massive systems, the initial period at which the binary will survive the MS without merging lies beyond the edge of our grid. These models may still merge when the star that leaves the main sequence swells up during H-shell burning. Since however, the post-MS phase of a star is at least an order shorter in duration than its MS lifetime, we do not expect models undergoing contact after their MS phase to contribute significantly to the observed population of contact binaries. We also report that none of our models evolve completely chemically homogeneously.  

Our models with $\textrm{q}_\textrm{i}=1$ are special in that, both stars have exactly the same mass. They hence evolve simultaneously and fill their Roche lobes at the same time. They do not exchange mass but keep expanding on their nuclear timescale until L2 overflow occurs. Hence the duration of the contact phase from these models is longer than their corresponding models with unequal initial mass ratios (cf.,  Fig.~\ref{scatter_contact_LMC}). The contribution of binaries with stellar masses that are exactly equal at birth is not well known, hence we reduce the importance of the $\textrm{q}_\textrm{i}=1$ models by our choice of the integration boundaries (cf., Section~\ref{distributions}) 

Generally we see that systems with the longest contact duration (those with the darkest shade of purple in Figs.~\ref{scatter_sub} and \ref{scatter_contact_LMC}) are concentrated at initial orbital periods at 0.8 days and total masses of 20\Msun (the lowest in our grid) across all initial mass ratios. The model with the longest contact duration in our LMC grid has P$_\textrm{i}=0.8$\,days, M$_\textrm{T,i}=20$\Msun and q$_\textrm{i}=1$ and spends about 10.1\,Myr in contact. The same model in the SMC grid spends about 10.7\,Myr in contact. 

\subsubsection{The impact of initial binary parameters on the overall grid population}

In Fig.~\ref{P_MT_initial} we fold the contact durations for all systems in each grid (Fig.~\ref{scatter_contact_LMC}) along with the birth weight of each system, to give us the likelihood of finding a contact binary in our synthetic population with a given initial period and mass ratio (cf., Eq.~\ref{eq:X_Pi} in Section~\ref{initial_hist}). The more purple the background color for each pixel is, the more likely that combination of P$_\textrm{i}$ and M$_\textrm{T,i}$ produces a contact binary. Between our LMC and SMC populations the main difference is the curves of `Initial L2 overflow' below which all models merge at the beginning of their binary evolution. We show these curves in Fig.~\ref{P_MT_initial} for the q$_\textrm{i}=1$ models. Owing to their higher metallicity, stars of the same initial mass are bigger in radii in the LMC compared to the SMC, due to which more binaries merge initially at larger separations in the LMC than the SMC. Apart from the shift in this boundary of `Initial L2 overflow' we find remarkably similar results between the SMC and LMC grids (see also our comparison between these grids in Fig.~\ref{age_hist} and \ref{frac_hist}).

The general trends already observed in the panels in Fig.~\ref{scatter_sub} are still visible. Contact binaries originate predominantly from the longer lived lower mass systems with short orbital periods. Models that have total masses of $20-24$\Msun dominate the population of contact binaries across all initial periods and mass ratios considered in our study. This is because along with their actual contact lifetimes, these systems are also weighted higher by the IMF, although it should be noted that we have used a log scale for the color bar in Fig.~\ref{P_MT_initial} and this may enhance features that could only be seen in a population that contains hundreds of contact binaries.

Systems which avoid merging during the MS evolution occur for a larger range of initial orbital periods and total masses in the SMC grid compared to the LMC, as their overall radii are smaller than their LMC counterparts. The large majority of these systems have initial orbital periods larger than 1.3 days for the SMC grid and 1.5 days for the LMC grid.  We also see from Fig.~\ref{scatter_sub} that the binaries which avoid merging on the MS only come in contact over a thermal timescale or do not come in contact at all. The implication for the observed contact binaries which are in contact over nuclear timescales thus is that, they  will merge before either of the two stars in the binary finishes core hydrogen burning.

\section{Comparison with observations}\label{results2}

\begin{figure*}
\centering
\begin{subfigure}{0.5\linewidth}
  \centering
  \includegraphics[width=\linewidth,left]{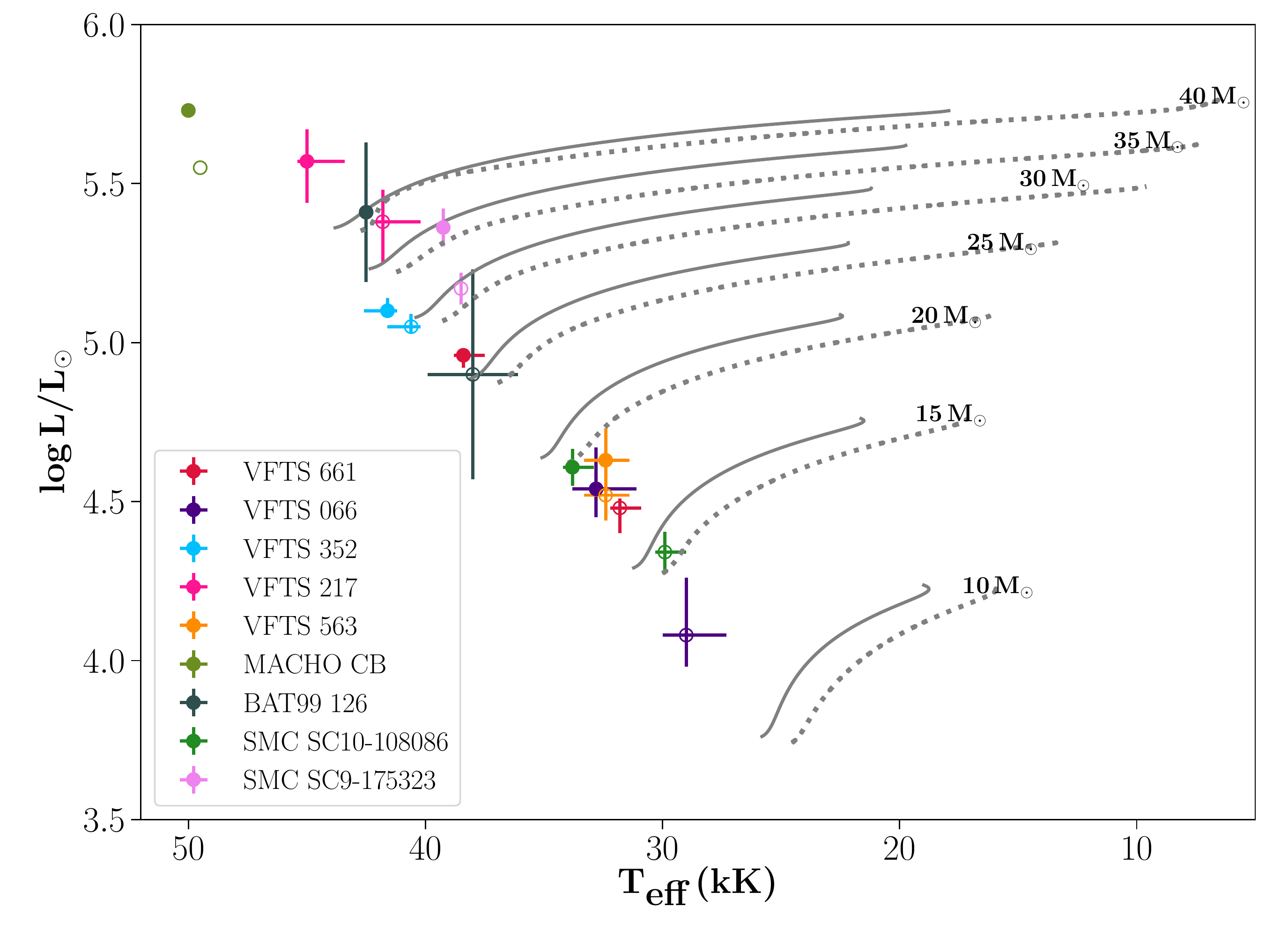}
\end{subfigure}%
\begin{subfigure}{0.5\linewidth}
  \centering
  \includegraphics[width=\linewidth]{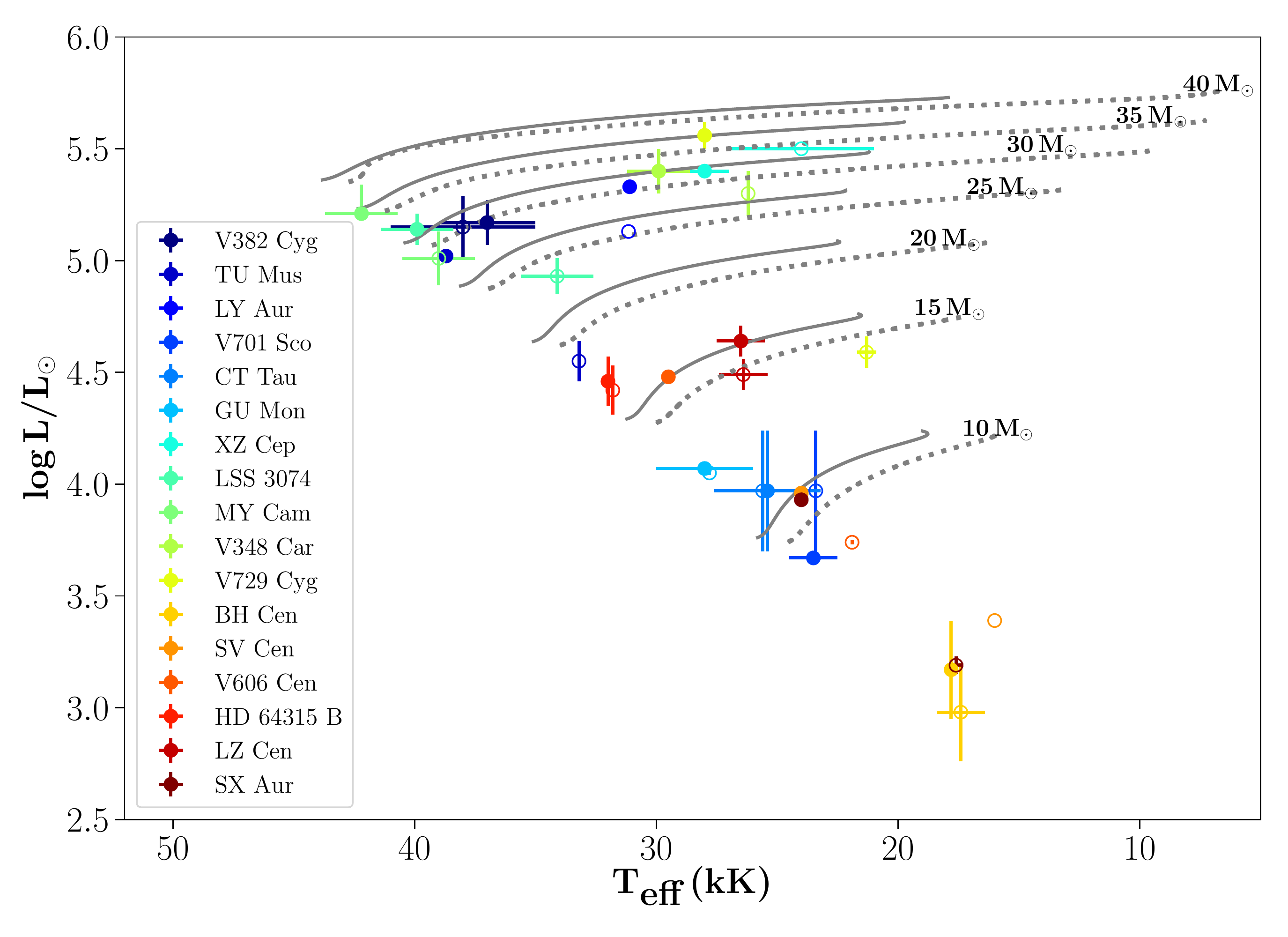}
\end{subfigure}
\caption{The Hertzsprung-Russel diagram of contact binaries from the Magellanic Clouds (left) and Milky Way (right). Where not directly reported for the system, we calculate the log\,L/L$_\odot$ with the T$_\textrm{eff}$ and R$_\textrm{mean}$ values in 
Tables~\ref{MC_contact} and \ref{gal_contact}. For reference, we also plot the evolutionary tracks of rotating (solid lines) and non-rotating (dotted lines) single-star models, with the LMC and solar metallicities, and with initial masses from 10...40\Msun. We compute these models until the end of their MS phase and  with the same physics assumptions as in Section~\ref{physics}. For the rotating models, we choose 
v$_\textrm{ini}=330\,$km\,s$^{-1}$, appropriate for the typical rotation rate expected for tidally locked stars in contact binaries.}
\label{HRD_contact}
\end{figure*}

\begin{table*}
\caption{Observed parameters for O\&B type massive contact systems in the LMC and SMC, including the orbital period (P$_\textrm{system}$), mass ratio q$_\textrm{system}$, dynamical mass (M$_{\star}$), mean radius (R$_\textrm{mean}$) and effective temperature T$_\textrm{eff}$  of individual components. We also provide the status known for the system: `C' for a confirmed contact or over-contact system, `NC' for a confirmed near-contact system and `C/NC' implies the status is not explicitly mentioned in the reference  or it could be in either configuration due to the error bars on its Roche-lobe filling factors (cf., Section~\ref{observed_sample} for more details). `P' and `S' stand for the primary (more massive) and secondary star of the system. The superscript `\textbf{T}' denotes the confirmed presence of a tertiary system. The system MACHO CB stands for MACHO*05:34:41.3$\pm$69:31:39. Error bars are absent for some primary T$_\textrm{eff}$ values as they were fixed to calculate the stellar parameters of the system.}
 \label{MC_contact}
 \begin{tabular}{llccccccccl}
  \hline
Contact system & Galaxy & P$_\textrm{system}$ (days) & q$_\textrm{system}$ (M$_\mathrm{S}/$M$_\mathrm{P}$) & M$_{\star}$\Msun& R$_\textrm{mean}$\,(R$_\odot$) & T$_\mathrm{eff}$\,(kK)  & Status &  Reference \\
  \hline
VFTS 661-P & LMC & 1.266 & 0.71$^{0.02}_{0.02}$ & 27.3$^{ 0.9}_{1.0}$ & 6.8$^{0.0}_{0.0}$ & 38.4$^{ 0.9}_{0.4}$ & NC & \citet{mahy2020b}\\
VFTS 661-S &       &    &     & 19.4$^{ 0.6}_{0.7}$ & 5.7$^{0.0}_{0.0}$ & 31.8$^{1.4}_{0.6}$ \\

VFTS 066-P & LMC & 1.141 & 0.52$^{0.05}_{0.05}$ & 13.0$^{7.0}_{5.0}$ &  5.8$^{0.5}_{0.8}$ & 32.8$^{1.7}_{1.0}$  & C/NC & \citet{mahy2020b}\\
VFTS 066-S &       &                      & & 6.6$^{3.5}_{2.8}$  & 4.4$^{0.4}_{0.8}$ & 29.0$^{1.0}_{1.2}$ &  & \\

VFTS 352-P & LMC & 1.124 & 0.98$^{0.02}_{0.02}$ & 25.6$^{1.7}_{1.4}$ & 6.8$^{0.1}_{0.2}$ & 41.6$^{0.4}_{1.0}$ & C &\citet{mahy2020b} \\
VFTS 352-S &       &                      & & 25.1$^{1.6}_{1.4}$ &  6.8$^{0.1}_{0.2}$ & 40.6$^{0.4}_{1.6}$ & \\

VFTS 217-P & LMC & 1.855 & 0.83$^{0.01}_{0.01}$ & 46.8$^{11.7}_{11.5}$ &  10.1$^{1.5}_{1.2}$ & 45.0$^{1.6}_{0.4}$  &  C/NC  &  \citet{mahy2020b}\\ 
VFTS 217-S &       &                      & & 38.9$^{9.7}_{9.7}$ &  9.4$^{1.4}_{1.0}$ & 41.8$^{1.7}_{0.6}$ \\ 

VFTS 563-P & LMC & 1.217 & 0.76$^{0.07}_{0.07}$ & 26.2$^{11.9}_{5.2}$ & 6.6$^{0.4}_{0.7}$ &  32.4$^{1.0}_{0.9}$  & C/NC & \citet{mahy2020b}\\
VFTS 563-S &       &                      & & 20.0$^{9.1}_{3.9}$ & 5.8$^{0.4}_{0.6}$ & 32.4$^{1.2}_{0.9}$ \\

MACHO CB-P & LMC & 1.400 & 0.64$^{0.01}_{0.01}$ & 41.0$^{1.2}_{1.2}$ &  9.6 & 50.0 & C/NC & \citet{ostrov2001} \\
MACHO CB-S &     &  & & 27.0$^{1.2}_{1.2}$ & 8.0 & 49.5  \\

BAT99 126-P & LMC & 1.550 & 0.41$^{0.08}_{0.08}$ & 36.5$^{1.2}_{1.2}$ &  9.4$^{1.8}_{1.8}$&  42.5 & NC & \citet{janssens2021}$^{\textbf{T}}$ \\
BAT99 126-S &     &  & & 15.0$^{2}_{2}$ & 6.7$^{1.7}_{1.7}$ & 38$^{1.9}_{1.9}$ & \\

OGLE SMC SC10-108086-P & SMC & 0.883 & 0.85$^{0.06}_{0.06}$ & 16.9$^{1.2}_{1.2}$ & 5.7$^{0.2}_{0.2}$  & 33.6$^{1.0}_{1.0}$ & C & \citet{hilditch2005}\\
OGLE SMC SC10-108086-S &      &      &   & 14.3$^{1.7}_{1.7}$ & 5.3$^{0.2}_{0.2}$ & 34.2$^{1.5}_{1.5}$  \\

OGLE SMC SC9-175323-P & SMC & 2.205 & 0.69$^{0.05}_{0.05}$ & 23.6$^{1.6}_{1.6}$ & 10.2$^{0.3}_{0.3}$  & 39.2 & NC  & \citet{harries2003}\\
OGLE SMC SC9-175323-S &      &      &   & 16.2$^{1.5}_{1.5}$ & 8.5$^{0.2}_{0.2}$ & 38.5 \\

  \hline
 \end{tabular}
\end{table*}

\begin{table*}
\caption{Observed parameters of early-type massive contact binaries in the MW. Table headings are the same as Table~\ref{MC_contact} except for the last column where we also include other papers which have studied a given system. Additionally in the last column, the superscript `\textbf{T}' denotes the confirmed/possible presence of a tertiary companion or system and `\textbf{A}' denotes the presence of an accretion disk. Where possible we report the dynamical mass of the components else we report the evolutionary or spectroscopic masses. Systems marked with a $\dagger$ symbol are much further evolved from ZAMS. Finally, error bars on all parameters are only mentioned up to the first significant digit.}
 \label{gal_contact}
 \centering
 \begin{tabular}{lccccccll}
  \hline
Contact system & P$_\textrm{system}$\,(days) & q$_\textrm{system}$ & M$_{\star}$\Msun & R$_\textrm{mean}$\,(R$_\odot$) & T$_\mathrm{eff}$\,(kK) &  Status & Reference & Previous works \\
  \hline
V382 Cyg-P & 1.885 & 0.73$^{0.01}_{0.01}$ & 26.1$^{0.4}_{0.4}$ & 9.4$^{0.2}_{0.2}$ & 37.0$^{2.0}_{2.0}$ & C/NC & \citet{martins2017} & \citet{harries1997a} \\
V382 Cyg-S &  &  & 19.0$^{0.3}_{0.3}$ & 8.7$^{0.2}_{0.2}$ & 38.0$^{3.0}_{3.0}$ & & & \citet{qian2007} \\

TU Mus-P & 1.387 & 0.65 & 16.7$^{ 0.4}_{0.4}$ & 7.2$^{0.5}_{0.5}$ & 38.7 &  C/NC & \citet{penny2008} & \citet{terrell2003} \\
TU Mus-S &  &  & 10.4$^{0.4}_{0.4}$ & 5.7$^{0.5}_{0.5}$ &  33.2 & & & \citet{qian2007}$^{\textbf{T}}$ \\

LY Aur-P $\dagger$& 4.002 & 0.55 & 25.5 & 16.1 & 31.1 & NC & \citet{mayer2013}$^{\textbf{T}}$ & \citet{stickland1994}\\
LY Aur-S &  &  & 14.0 & 12.6 & 31.1 &   &  & \citet{zhao2014}$^{\textbf{T}}$\\

V701 Sco-P & 0.762 & 0.99 & 9.8$^{0.2}_{0.2}$ & 4.1$^{0.4}_{0.4}$ & 23.5$^{1.0}_{1.0}$ & C & \citet{yang2019} &  \citet{hilditch1987} \\
V701 Sco-S  &  &  & 9.7$^{0.2}_{0.2}$ & 4.1$^{0.2}_{0.2}$ & 23.4$^{0.1}_{0.1}$ & &  & \citet{qian2006a}$^{\textbf{T}}$\\

CT Tau-P & 0.666 & 0.98 & 14.2$^{3.3}_{3.3}$ & 4.9$^{0.4}_{0.4}$ & 25.4$^{2.2}_{2.2}$ & C & \citet{yang2019} & \citet{plewa1993}  \\
CT Tau-S  &  &  & 14.0$^{3.4}_{3.4}$ & 4.9$^{0.2}_{0.2}$ & 25.6$^{0.2}_{0.2}$ & & \\

GU Mon-P & 0.896 & 0.97 & 8.8$^{0.1}_{0.1}$ & 4.6$^{0.25}_{0.25}$ & 28.0$^{2.0}_{2.0}$ & C & \citet{yang2019} & \\
GU Mon-S &  &  & 8.6$^{0.1}_{0.1}$& 4.6$^{0.2}_{0.2}$ & 27.8$^{0.07}_{0.07}$ &  & \\

XZ Cep-P $\dagger$ & 5.097 & 0.50$^{0.01}_{0.01}$ & 18.7$^{1.3}_{1.3}$ & 14.2$^{0.1}_{0.1}$ & 28.0$^{1.0}_{1.0}$ & NC & \citet{martins2017} & \citet{harries1997a} \\
XZ Cep-S &  &  & 9.3$^{0.5}_{0.5}$ & 14.2$^{0.1}_{0.1}$ & 24.0$^{3.0}_{3.0}$ &  & \\

LSS 3074-P & 2.185 & 0.86 & 14.6$^{2.1}_{2.1}$ & 7.5$^{0.6}_{0.6}$ & 39.9$^{1.5}_{1.5}$ & C/NC & \citet{raucq2017} & \\
LSS 3074-S &  &  & 17.2$^{3.0}_{3.0}$ & 8.2$^{0.7}_{0.7}$ & 34.1$^{1.5}_{1.5}$ & & \\

MY Cam-P & 1.175 &  0.84$^{0.03}_{0.03}$& 37.7$^{1.6}_{1.6}$ & 7.6$^{0.1}_{0.1}$ & 42.2$^{1.5}_{1.5}$ & C/NC &\citet{lorenzo2014} & \\
MY Cam-S &  &  & 31.6$^{1.4}_{1.4}$ & 7.0$^{0.1}_{0.1}$ & 39.0$^{1.5}_{1.5}$  & & \\

V348 Car-P $\dagger$& 5.600 & 0.95$^{0.05}_{0.05}$ &  35$^{1.0}_{1.0}$ & --  & 29.9$^{1.3}_{1.3}$ & NC & \citet{hilditch1985} & \\
V348 Car-S &  &  & 35$^{1.0}_{1.0}$ & -- & 26.2 &  & \\

V729 Cyg-P $\dagger$ & 6.597 & 0.29$^{0.04}_{0.04}$ & 31.6$^{2.9}_{2.9}$ & 25.6$^{1.1}_{1.1}$ & 28.0 & C & \citet{yasaroy2014} &   \citet{linder2009}$^{\textbf{T}}$\\
V729 Cyg-S &  &  & 8.8$^{3}_{3}$ & 14.5$^{1.0}_{1.0}$ & 21.3$^{0.4}_{0.4}$ & &  \citet{kennedy2010}$^{\textbf{T}}$ \\

BH Cen-P & 0.792 & 0.84--0.885 & 9.4$^{5.4}_{5.4}$& 4.0$^{0.7}_{0.7}$ & 17.8 & C/NC & \citet{leung1984} &   \citet{qian2006a}$^{\textbf{T}}$ \\
BH Cen-S &  &  & 7.9$^{5.4}_{5.4}$ & 3.7$^{0.7}_{0.7}$ & 17.4$^{1.0}_{1.0}$ &  & \citet{zhao2018}$^{\textbf{T}}$ \\

SV Cen-P & 1.658 & 0.80 & 7.7 & 7.3 & 24.0 & NC & \citet{linnel1991}$^{\textbf{A}}$ & \cite{drechsel1982}  \\
SV Cen-S &  &  & 9.6 & 7.8 & 16.0 & &  \citet{shematovich2017}$^{\textbf{A}}$ \\

V606 Cen-P & 1.490 & 0.53$^{0.02}_{0.02}$ & 14.3$^{0.41}_{0.41}$ & 6.8$^{0.06}_{0.06}$ &  29.5 & C/NC & \citet{lorenz1999} & \\
V606 Cen-S &  &  & 8.0$^{0.24}_{0.24}$ & 5.13$^{0.5}_{0.5}$ & 21.9 \\

HD 64315 B-P & 1.019 & 1.00$^{0.06}_{0.06}$ & 14.6$^{2.3}_{2.3}$ & 5.5$^{0.5}_{0.5}$ &  32.0 & C & \citet{lorenzo2017}$^{\textbf{T}}$ \\
HD 64315 B-S &  &  & 14.6$^{2.3}_{2.3}$ & 5.3$^{0.5}_{0.5}$ & 31.8  \\

LZ Cen-P & 2.75 & 0.92$^{0.07}_{0.07}$ & 13.5$^{1.4}_{1.4}$ & 9.1$^{0.3}_{0.3}$  &  26.5$^{1.0}_{1.0}$  & NC & \citet{vaz1995} \\
LZ Cen-S &  &  & 12.5$^{1.3}_{1.3}$ & 8.4$^{0.3}_{0.3}$  & 26.4$^{1.0}_{1.0}$  \\

SX Aur-P &1.21 & 0.61 & 11.3$^{0.2}_{0.2}$ & 5.3   &  
24.0$^{0.3}_{0.3}$  & NC & \citet{ozturk2014} \\
SX Aur-S &  &  & 6.9$^{0.1}_{0.1}$ & 4.2 & 17.6$^{0.3}_{0.3}$ \\
 \hline
 \end{tabular}
\end{table*}

\subsection{The observed sample}
\label{observed_sample}

We have collected the data of 26 spectroscopic massive contact binaries from the Magellanic Clouds (Table~\ref{MC_contact}) and the Galaxy (Table~\ref{gal_contact}), for which the orbital and stellar parameters are available. These include 2 systems from the SMC, 7 from the LMC and 17 from the MW. The majority of these are O+O contact binaries, with a few B+B and O+B systems. Most of these systems are located close to ZAMS and a few are located much further along the MS (Fig.~\ref{HRD_contact}). A few of the Galactic systems have tertiary companions (or systems) reported as well. However, we do not model the effect of tertiary systems on the evolution of contact binaries as this is beyond the scope of our work. 

Of the 7 O-type spectroscopic contact binaries in the LMC, 4 are from the VFTS-TMBM sample. This is because the VFTS sample is largely biased (and complete) in O stars \citep{sana2013, almeida2017}, which in turn may be because 30~Dor itself has a larger concentration of very massive stars than the average LMC population \citep{schneider2018a, schneider2018b}. In addition to the VFTS sample, 29 `extremely-blue' contact binaries have been identified from the MAssive Compact Halo Objects (MACHO) survey data \citep{rucinski1999} , all located close to the Zero Age Main Sequence (ZAMS), with periods of 0.45-1.3\,days and a few outliers with periods up to 3.5\,days. From the Optical Gravitational Lensing Experiment (OGLE)-III catalogue, 28 early-type  contact binaries with periods between 0.5 and 1.6\,days have been reported \citep{pawlak2016}. 
There is an overlap between the two data sets and roughly, we estimate there to be 30 unique O \& B-type contact binaries together from the MACHO and OGLE database for the LMC. Unfortunately, since these are photometric binaries, the stellar parameters of individual systems are not available. 

Most contact systems are eclipsing binaries that are distinguished by their smooth, nearly-sinusoidal light curves which do not show plateaus between eclipses \citep{lorenzo2017}, indicating that both stars have near-equal brightness and show signs of tidal distortion \citep{hilditch1998}. By modelling the light curve and with the radial velocity measurements, one can discern the geometry of the binary and the degree by which the stars overflow their respective Roche lobe volumes-- their `fill-out factors'. 

It is in the uncertainty of these fill-out factors that the exact nature of a system becomes unclear; whether it is truly a contact binary, or if it is simply approaching contact. The main contributor to this uncertainty is the inclination of the system-- the further away from 90$^{\circ}$ the inclination is, the larger the error bars are on the measurements of the stellar parameters.

The fill-out factors themselves are reported differently, depending on how they are calculated. One method to calculate the fill-out factor ($f_{\Omega}$) is by comparing the difference between the surface potential ($\Omega$) of the stars (assumed to be identical for either star) with the potentials at the inner and outer Lagrangian points of the system \citep{mochnacki1972}. A value of $f_{\Omega}=0$ indicates a system exactly in contact and $f_{\Omega}>0$ indicates an over-contact system \citep{lorenz1999}. By this definition, V606 Cen which has  $f_{\Omega}=0.01-0.04$ is considered to be marginally in contact \citep{lorenz1999}, V729 Cyg with $f_{\Omega}=0.17-0.22$ as a binary in moderate contact \citep{yasaroy2014}, and V701 Sco, GU Mon and CT Tau with $f_{\Omega}=0.55,0.70$ and 0.99 respectively, as systems in deep contact \citep{yang2019}.

A second approach, is to find the (over)filling factor f$_\textrm{L}$, by comparing the average volume of each star to its respective Roche lobe volume, i.e., $f_\textrm{L}=(R_\textrm{mean}/R_\textrm{RL})^{3}$. A value of $f_\textrm{L}=1$ thus indicates a system exactly in contact. One of the hottest and most massive contact binaries known, VFTS 352, has $f_\textrm{L}=1.29$ for both stars \citep{almeida2015}, thus assigning its status as a binary in deep contact.  Other systems can also be classified as over-contact binaries based on this definition, such as  V382 Cyg  ($f_\textrm{L}=1.1$ for both stars), OGLE SMC-SC10 108086 ($f_\textrm{L}=1.7$ for both stars)  and TU Mus ($f_\textrm{L}=1.2, 1.3$) (these $f_\textrm{L}$ values are as reported by \citealp{almeida2015}). Despite  $f_\textrm{L}$ being higher than 1 for V382 Cyg, \citet{martins2017} who derived the parameters and geometry of V382 Cyg, report it as a system in which both stars are barely filling their Roche lobes. \citet{penny2008} who studied TU Mus, indicate it as a system that is either approaching contact or is in marginal contact. The status of another Galactic contact binary, LY Aur, is also uncertain, with \citet{mayer2013} indicating that it may even be a semi detached system. SX Aur is reported as a near-contact binary by \citet{ozturk2014} who calculated a value of  $f_\textrm{L}=1.02$ for both its stars. Hence although the system LSS 3074 is reported as an over-contact binary by \citet{raucq2017}, given its $f_\textrm{L}=1.008\pm=0.01$,  we suspect this may also be a near-contact binary. Other reported near-contact binaries are: OGLE SMC-SC9 175323 \citep{harries1997a}, LZ Cen \citep{vaz1995}, XZ Cep \citep{martins2017}, V348 Car \citep{hilditch1987} and BAT99 126 \citep{janssens2021}. 

For some systems, the fill-out factors are not mentioned in the works that study them; these are MACHO*05:34:41.3$\pm$69:31:39 \citep{ostrov2001}, VFTS\,066 \citep{mahy2020a,mahy2020b}, HD 64315 B \citep{lorenzo2014} and MY Cam \citep{lorenzo2014}. As their degree of contact is not reported, we assign these systems a `C/NC' status (contact/near-contact) in Tables~\ref{MC_contact} and \ref{gal_contact}, along with V382 Cyg, TU Mus and LY Aur in the Milky Way and, VFTS\,217 and VFTS\,563 in the LMC, which
are reported as systems with ``uncertain configurations'' \citep{mahy2020b}, but which have large enough errors on their R$_\textrm{mean}$/R$_\textrm{L}$ values to allow them to be contact systems as well. The same holds for VFTS\,066 which is reported as a contact binary from its light curve, but again has a large range in R$_\textrm{mean}$/R$_\textrm{L}$ to also allow it to be a near-contact binary \citep{mahy2020a, mahy2020b}. The fill-out factors of contact binaries from the OGLE+MACHO database are also unknown.

From our literature survey thus, the list of confirmed massive over-contact binaries include: V701 Sco, CT Tau, GU Mon, V729 Cyg, VFTS 352 and OGLE SMC-SC 108086. Systems reported with equal masses ($\textrm{q} \approx 1$) are V701 Sco, CT Tau, GU Mon, HD 64315, V348 Car and VFTS 352,  while systems with near-equal masses within error bars ($\textrm{q}\gtrapprox0.9$) are OGLE SMC SC10-108086, BH Cen and LZ Cen.

\begin{figure*}
\centering
\begin{subfigure}{0.425\linewidth}
\centering
  \includegraphics[width=\linewidth,left]{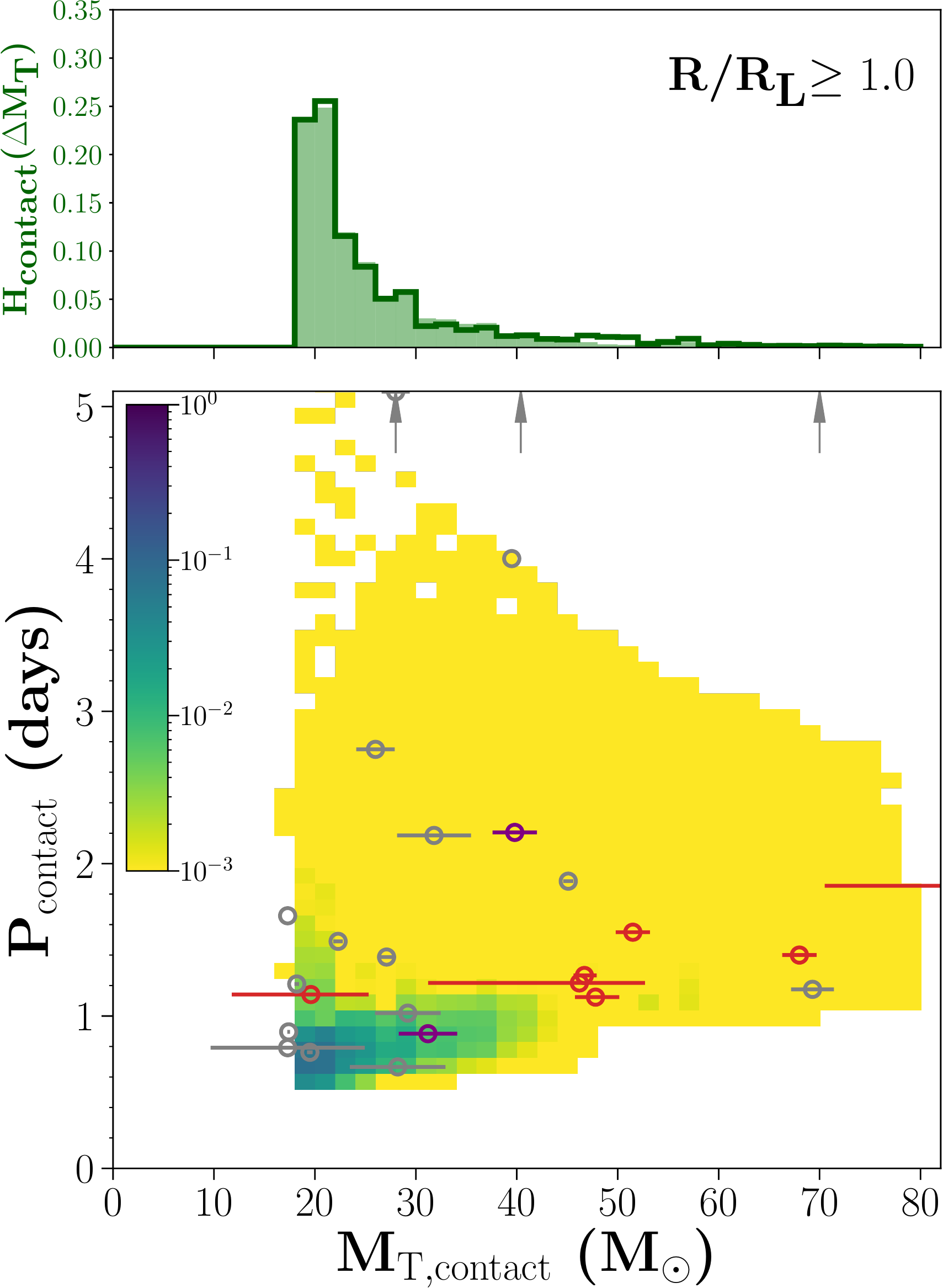}
\end{subfigure}%
\begin{subfigure}{0.55\linewidth}
  \centering
  \includegraphics[width=\linewidth,clip]{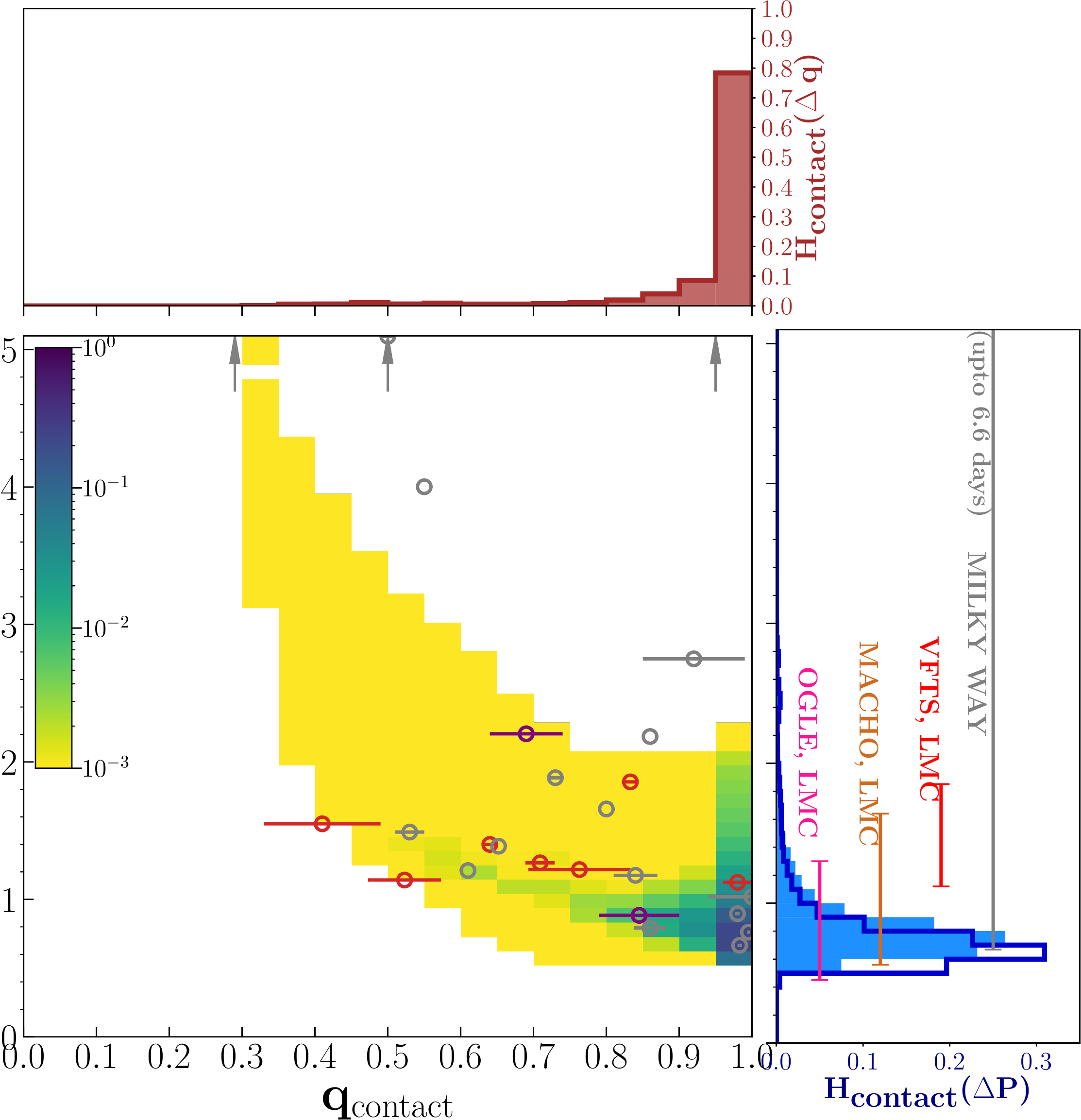}
\end{subfigure}
\par\bigskip
\centering
\begin{subfigure}{0.425\linewidth}
\centering
  \includegraphics[width=\linewidth,left]{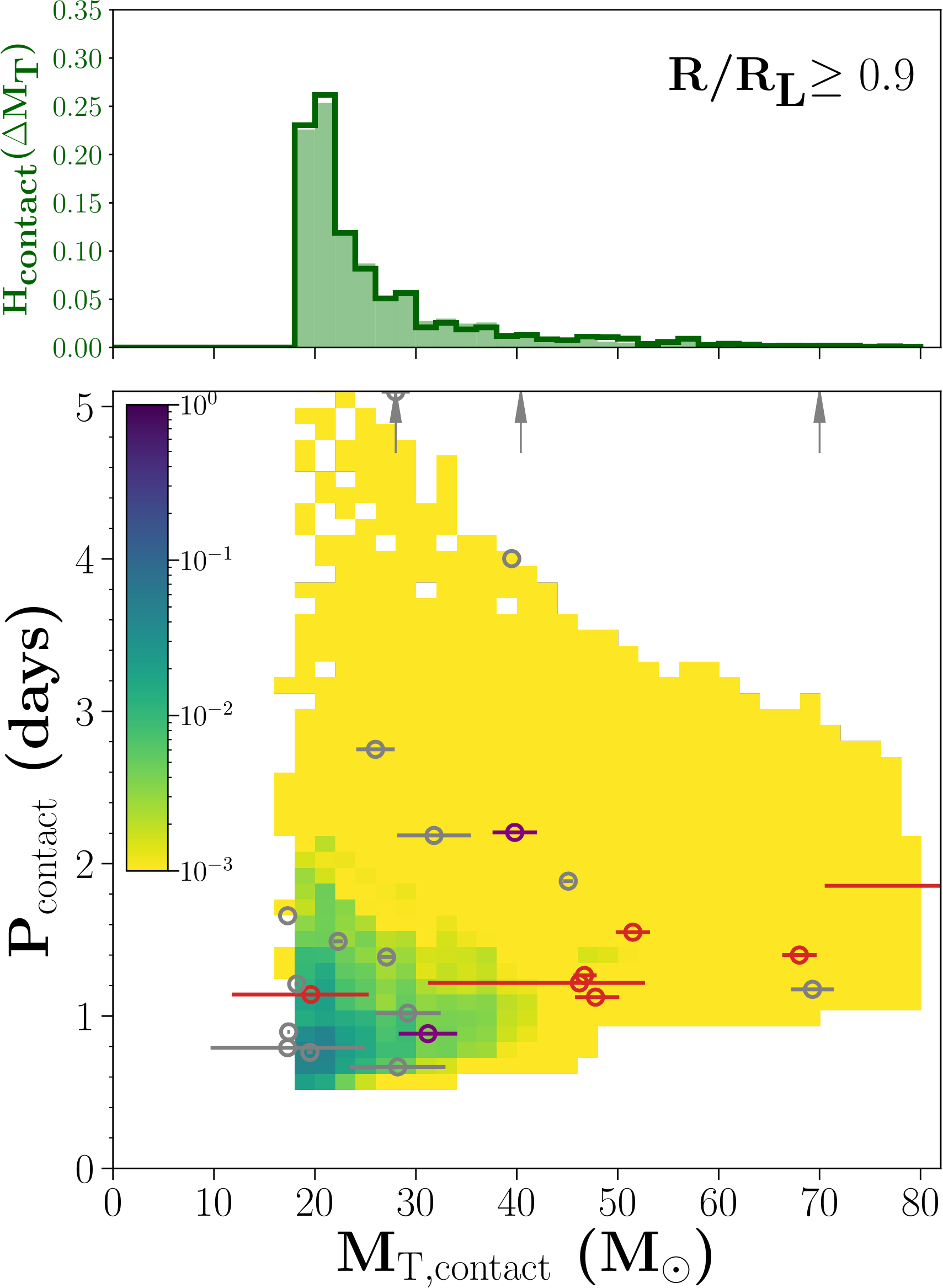}
\end{subfigure}%
\begin{subfigure}{0.55\linewidth}
  \centering
  \includegraphics[width=\linewidth,clip]{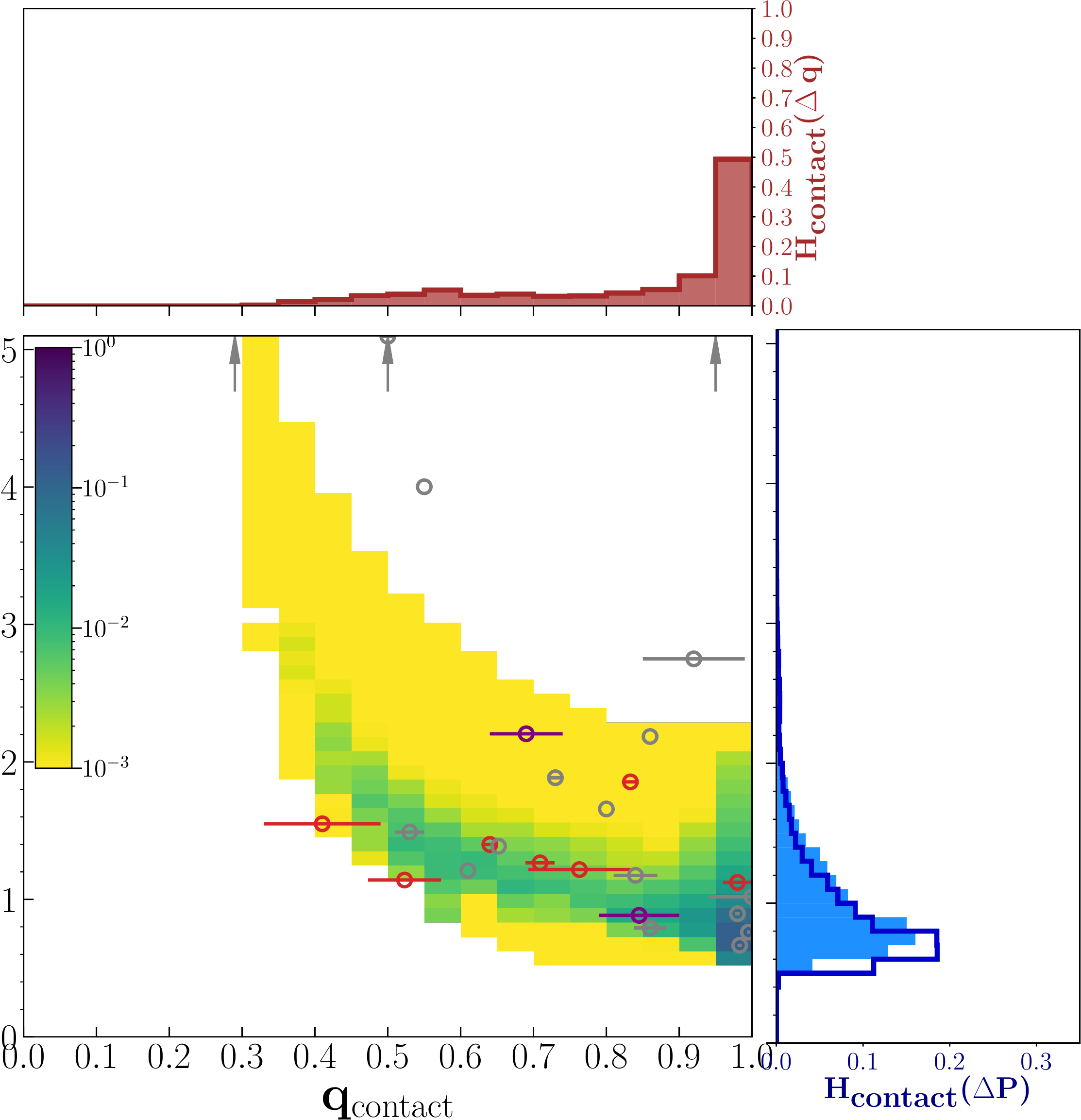}
\end{subfigure}
\caption{Probability distribution of observed period and total mass (P$_\textrm{contact}$--M$_\textrm{T,contact}$) and of observed period and mass ratio (P$_\textrm{contact}$--q$_\textrm{T,contact}$), for our LMC population of contact binaries (top plots) and relaxed-contact binaries.  The background colour represents the probability of finding a system with a given combination of 
orbital parameters (cf., Section~\ref{contact_dist}). The 1D shaded histograms on top and to the right are the corresponding projections of the parameters plotted in the 2D histograms. Over-plotted in solid dark lines in the 1D histograms are the corresponding distributions from the SMC grid.  Observed contact systems are also overplotted; red markers for LMC, purple marker for SMC (Table~\ref{MC_contact}) and grey markers for MW systems (Table~\ref{gal_contact}). The observed period range of contact binaries from the OGLE, MACHO,
and VFTS samples are marked in the P$_\textrm{contact}$ histogram (blue, top panel). The grey arrows in the plots indicate the position of the two MW binaries with P$_\textrm{contact}>$5\,days.}
\label{LMC_contact}
\end{figure*}

\subsection{Prediction for orbital parameters in the observed population of massive contact binaries}

In this section we present the results of our 2D probability distributions for the currently observed  parameters of contact binaries for the LMC and SMC, assuming that star formation occurs at a constant rate.
Figure~\ref{LMC_contact} shows the probability of finding a contact system in the LMC with a particular combination of observed period, P$_\textrm{contact}$ and total mass, M$_\textrm{T,contact}$, or with a particular period and mass ratio q$_\textrm{contact}$ (cf., Section~\ref{contact_dist} for the construction of these distributions).  The upper two plots consider only those models which are strictly in contact 
(where both stars have R/R$_\textrm{L}\geq1$), while the lower two plots include all models in 
relaxed contact, where both stars have R/R$_\textrm{L}\geq0.9$ (cf., Section~\ref{examples}).  We also compare our predicted distributions with the available data of contact binaries in the Magellanic Clouds and the Galaxy (Tables~\ref{MC_contact} and ~\ref{gal_contact}). The distributions of the above parameters for the SMC are very similar to those of the LMC, as we can anticipate from the 2D histograms in Fig.~\ref{P_MT_initial}. Therefore the discussion of our results for the LMC also applies to the SMC. We also include the Galactic contact binaries in our discussion for the same reason, although of course, a dedicated solar-metallicity grid would be desirable for a proper analysis of these binaries.

As the contact duration of a binary is strongly dependent on its initial mass, we need to adapt our synthetic population to suit the observed populations which are dominated by very massive systems. Hence, we also compute P$_\textrm{contact}$-q$_\textrm{contact}$ distributions for the LMC and SMC grids by applying two different initial mass cuts on the synthetic contact binary population, i.e., with M$_\textrm{T,i}\geq30\Msun$ and M$_\textrm{T,i}\geq50\Msun$. These are shown in Fig.~\ref{masscut_contact}.

In both the top and bottom panels of Fig.~\ref{LMC_contact}, we see a tight correlation in the predicted distribution between M$_\textrm{T,contact}$ and P$_\textrm{contact}$; the more massive a contact binary is, the larger its current orbital period will be. Another (anti) correlation can also be seen between the P$_\textrm{contact}$ and q$_\textrm{contact}$; wider systems are expected to have smaller mass ratios. 

The average orbital period expected for (over-)contact binaries is less than 1\,day in the LMC and SMC, with a maximum of $\approx 1.5$\,days for the lowest mass contact binaries. The mass ratio distribution from this population predicts that nearly 90\% of over-contact systems must have mass ratios $\geq 0.9$. Not surprisingly, the most likely total mass of contact systems is close to the lower total mass boundary of our grids, due to the IMF and lifetime effects and also because the contact duration increases for smaller initial mass and initial period. The most likely orbital periods are found strongly skewed towards the smallest period values with P$_\textrm{contact}$ in the range 0.6$\dots$1\,day for both metallicities, and a peak slightly lower for the SMC (at 0.7\,days) than the LMC (at 0.8\,days).

We see that the over-contact SMC system OGLE SMC SC10-108086, falls well within the probable parameter space of over-contact binaries.  Among the LMC contact binaries, the lowest mass system VFTS\,066, has an orbital period which is about 30\% larger than the most likely period at this total mass. While this could be significant, a comparison with the mass ratio distribution (right panel) shows that for its mass ratio of about 0.52, a larger period than the most likely period for this total mass is actually expected. The other five systems, lie in fact all very close to the most likely orbital period for a given system mass, which is especially clarified by Fig.~\ref{masscut_contact}. This even seems to hold for VFTS\,217, whose mass exceeds that of our most massive models; however, this system is most likely to be a near-contact binary. We conclude that while statistics is sparse, our models reproduce the observed periods of the Magellanic Cloud contact binaries  well. We also note a strong agreement between the Galactic over-contact binaries and our predictions.

Looking at the mass ratios of the observed contact binaries in  Fig.~\ref{LMC_contact}, we see that indeed the confirmed over-contact binaries (VFTS\,352 and SMC SC10-108086) in the Magellanic Clouds have q$_\textrm{contact}=0.9-1$ within their error bars, as do the three confirmed over-contact systems in the Galaxy, along with two others whose status is unclear. However, given how skewed the mass ratio distribution is towards equal masses, we would expect the majority of observed systems to cluster in the  q$_\textrm{contact}\geq 0.9-1$ space. This mismatch is curious, since, considering the mass dependence of the period distribution, most of the observed contact binaries follow the predicted tight correlation between orbital period and mass ratio rather closely. This is especially evident in Fig.~\ref{LMC_contact} systems with M$_\textrm{T,contact}\leq40$\Msun in the systems from the Magellanic Clouds as well as from the Galaxy, and in Fig.~\ref{masscut_contact} for the more massive ones.

A more appropriate comparison for the observed systems would be with the predictions from our mixed population of near-contact and over-contact binaries, shown in the bottom panels of Fig.~\ref{LMC_contact}.  By including the near-contact binaries (those in which both stars have $ 0.9 \leq \textrm{R}/\textrm{R}_\textrm{RL} < 1$), the range of q$_\textrm{contact}$ spreads down to 0.4 and that of P$_\textrm{contact}$ goes up to 2\,days. The majority of the observed contact binaries are well covered in this P$_\textrm{contact}$-q$_\textrm{contact}$ space. The near-contact binaries OGLE SMC SC9-175323 and VFTS\,217, however, still remain outliers in this space, along with some of the Galactic systems. Hence while the likelihood for finding  unequal-mass contact/near-contact binaries increases considerably, we see that, while the cut at 90\% of the Roche-lobe radius is somewhat arbitrary, this effect may alone be insufficient to explain the observed q-distribution.

From the M$_\textrm{T,contact}$-P$_\textrm{contact}$ diagram in Fig.~\ref{LMC_contact}, we see that many of the observed systems are heavier than what we expect from our synthetic population predictions.  We attribute this to the fact that TMBM focuses on O-star binaries \citep{sana2013, almeida2017}, while in our grid, many models belong to the lower-mass regime of early B stars. Hence another relevant factor to consider while comparing with observations, is the high mass bias in the observed sample. A comparison of Figs.~\ref{LMC_contact} and \ref{masscut_contact} shows that the predicted preference
for $\textrm{q}_\textrm{contact}\geq0.9$ decreases significantly for higher system masses--the preference for high mass ratios (q$_\textrm{contact}\geq 0.9$) is reduced from $\sim$90\% for our complete grid to $\approx60$\% for systems with initial masses above $50\Msun$.  We also note that whereas the distributions shown in Fig.~\ref{LMC_contact} show very little metallicity dependence, this is different in  Fig.~\ref{masscut_contact} where the predicted LMC period distributions appear to be shifted to significantly larger periods, compared to that of the SMC for the highest mass cut as seen in  Fig.~\ref{masscut_contact}.

While we did not model binaries with Galactic or solar metallicity, the similarity of our SMC and LMC results for the lower half of the considered mass range suggests that metallicity effects do not play a major role for system masses below $\sim45$\Msun. However, for more massive systems, differences become more prominent (cf., Fig.~\ref{masscut_contact}), and could even be larger when compared to the rather metal-rich MW.

While the LMC and SMC spectroscopic contact binaries have periods larger than 1.1\,days our predicted period for contact binaries can be as low as 0.6\,days. In this regard, we also consider the orbital period range of massive contact binaries from the MACHO \citep{rucinski1999} and OGLE-III \citep{pawlak2016} surveys, which also include near-contact binaries. Almost half of the massive contact binaries in these samples have periods $<=1\,$day, which agrees with the predictions of our distributions. It is important to note that the smaller periods in these samples compared to the VFTS-TMBM sample may reflect the fact that the OGLE binaries are not restricted to O\,stars alone, and could therefore mirror the early B\,star systems in our model grid. 

We see that, in comparison with the Magellanic Cloud systems the MW contact binaries are much more scattered in the observed parameter space. In particular, we find 6 contact binaries with orbital periods in excess of 2\,days (with a maximum value of 6.6\,days), whereas the largest period in the Magellanic Cloud systems is nearly 1.89\,days. The more extreme outliers in mass ratio and period in the MW sample, such as LY Aur, XZ Cep, V348 Car and V729 Cyg, are much further evolved from the ZAMS compared to those with periods close to or less than 2\,days (Fig.~\ref{HRD_contact}). Notably, three of these outliers except for XZ Cep, have a total mass of more than 40\,M$_{\odot}$. This may be because XZ Cep, although stated as a contact binary in \citet{martins2017}, is in fact a semi-detached system (\citealp{harries1997a}, private communication, Laurent Mahy).  

We speculate about possible reasons for this in Section~\ref{discussion}. On the other hand, the observed MW systems with orbital periods below $\sim$1.5\,days  follow the predicted distributions quite closely, including a concentration of four of them with mass ratios of almost one.

From the distributions in Figs.~\ref{scatter_sub} and \ref{P_MT_initial}, and in more detail from Fig.~\ref{scatter_contact_LMC}, we can obtain an idea about the completeness of our synthetic contact binary population. We see that in Fig.~\ref{scatter_contact_LMC}, towards the shortest orbital periods, our grid is certainly complete. For larger periods, we see that the darker purple shading remains restricted to initial orbital periods below $\sim1.5$\,days, while  our upper period limit is 2\,days.  Only for  q$_\textrm{i}=1$ we see contact times exceeding 1\,Myr at initial periods of 2\,days. However, at q$_\textrm{i}=1$, the nuclear timescale contact regime is limited by the maximum orbital period for Case A mass transfer, and below this limit the contact duration decreases stronger than linear with 
increasing period, reflecting the stellar radius evolution. Therefore, while we are missing  a fraction of contact binaries starting with q$_{\rm i}\simeq 1$, this fraction is quite small.  On the other hand, we may also be overestimating the number of contact binaries with q$_{\rm i}\simeq 1$, since our q$_{\rm i}\,= 1$ models have a statistical weight in our plots represented by the interval
$[1,0.975]$, but perhaps the contact timescales depicted in the q$_{\rm i}\,=1$ plot of Fig.~\ref{scatter_contact_LMC} are representative of a smaller q$_{\rm i}$-interval, which may explain the apparent overabundance of symmetric contact binaries in our synthetic population.

Concerning contact binaries that may originate from smaller initial mass ratios than the lowest value in our grid, 
 Fig.~\ref{scatter_contact_LMC} shows that the dark purple island indicating the longest contact times becomes much paler towards
the lowest considered initial mass ratios. We may miss a fraction of contact systems which originate from more extreme initial mass ratios than $\rm q_{\rm i} = 0.6$, but expect their integrated contact durations to be relatively small, compared to most other contact binaries in the population we model. We also see in Fig.~\ref{scatter_contact_LMC} that with initial mass ratios further from one, the plots contain increasingly large areas in which the MESA models did not converge. Again, we expect the contact durations of each of those models to be relatively small.  

There may, however, be local regions of the predicted parameter space in our grid is incomplete, either due to numerical instability or the extent of the initial conditions. This concerns in particular the models that fail to converge with high total masses and initial mass ratios further from one, as shown in the panels of q$_\textrm{i}=0.6$ and 0.7 of Fig.~\ref{scatter_contact_LMC}. The missing models in these regions may be the reason why the largest
predicted orbital periods of contact binaries with $\rm M_{\rm T,i} \geq 50 M_{\odot}$, in the top-right panel of Fig.~\ref{masscut_contact}, is about 3.5\,days. This is smaller than the largest orbital period predicted from our lower-mass contact binary models of about 5\,days, as seen in the top-right panel of Fig.~\ref{LMC_contact}, which conflicts with 
the expectation that more massive contact binaries would generally have longer orbital periods for the same mass ratio.  It is also worth noting systems with $\rm q_{\rm i} = 1.0$ could also contribute to even longer orbital periods, however, the probability of find binary systems that begin their evolution with with components of exactly the same mass is hard to estimate.

Overall, we argue that our coverage of contact binaries in initial period and mass ratio is sufficient to include the majority of the expected massive contact binary populations with SMC and LMC composition, for total system masses below $80\Msun$.

\subsection{The number of O-type contact systems in the Magellanic Clouds}
\label{results3}

We estimate the number of contact binaries in 30~Dor using our distributions, and thereby their numbers in the LMC and SMC.  We only calculate the number of O+O type systems in 30~Dor as the VFTS-TMBM data with which we compare our results,  is designed to have minimal biases for O-type stars, i.e., those with masses $\geq15$\Msun \citep{evans2011, sana2013}.

The total number of O stars in the VFTS sample is N$_\textrm{O-stars}=360$ and the binaries among them span a range in period of 1.1$\dots$3162\,days, and in mass ratio of 0.1$\dots$1.0 \citep{sana2013}. The maximum mass of an early O-type star in the sample is 90\Msun (VFTS~061, \citealp{ramirez2017}), thus lending us a mass range of $15-90$\Msun for O-type stars in the VFTS sample. The intrinsic binary fraction (f$_\textrm{bin}$) inferred for O-type stars in 30~Dor is  0.51$\pm$0.04  \citep{sana2013}, thus lending us a range in f$_\textrm{bin}$ of $\approx 0.47-0.55$. The orbital period range of the VFTS binaries is from 1.1\,days to to 3162\,days, while that of the mass ratio is from 0.1 to 1.0.

As we are interested only in O+O-type binaries, we select only those models from our grid which have both M$_\textrm{1,i}$ and M$_\textrm{2,i}$ as 15\Msun.  The fraction of O+O-type contact systems among MS binaries from our grid yields $\textbf{f}_\textrm{contact/MS}=0.31$ (Eq.~\ref{eq:f_contact_MS}) for the strict definition of contact and $\textbf{f}_\textrm{contact/MS}=0.50$ for the relaxed contact definition. 

To be consistent with our calculation of probability distributions (as in Section~\ref{distributions}), we  assume a Salpeter distribution for the IMF and a flat distribution in log\,period and also in mass ratio, to find the overlap between the initial parameter space coverage of our grid and that of the VFTS binaries. We calculate this by finding the fraction of the total birth weight of our grid, W$_\textrm{grid}$, compared to that of the VFTS sample W$_\textrm{VFTS}$. This fraction is  W$_\textrm{grid}$/W$_\textrm{VFTS}=0.03$. The number of contact binaries in 30~Dor using the VFTS parameters  $N_\textrm{contact, VFTS}$ can be estimated as:

\begin{equation}
\begin{split}
N_\textrm{contact, VFTS} \sim \left ( \frac{\textbf{f}_\textrm{contact/MS}}{0.31...0.50} \right) \times \left ( \frac{N_\textrm{O-stars}} {360} \right) \times \left ( \frac{f_\textrm{bin}} {0.47...0.55} \right) \\ 
\times \left ( \frac{\textrm{W}_\textrm{grid}/\textrm{W}_\textrm{VFTS}} {0.03} \right).
\end{split}
\end{equation}

We expect a maximum of 2 O-type over-contact binaries or 3 relaxed-contact systems  (comprising both the near and over-contact binaries) in 30~Dor. As there are currently 1 confirmed over-contact system and 3 contact/near-contact O-type binaries in the VFTS-TMBM sample (see systems marked as `C/NC' in Table~\ref{MC_contact}), our numbers are in close agreement with the observations. 

Since 30~Dor hosts about $\approx25$\% of the total number of O\,stars in the LMC (Crowther, P., private comm.), we can expect up to $N_\textrm{contact, LMC}\approx8$ O-type over-contact and up to 12 relaxed-contact (near + over-contact) binaries in the LMC, assuming like 30~Dor that only $\approx 50$\% of O-stars are in binaries. If we instead assume that the binary fraction is 100\% in the LMC as is the case for O-stars in the Milky Way (e.g., \citealp{sana2014}), our predicted numbers go up to 13 over-contact or 21 relaxed O-type contact binaries in the LMC. The total number of observed massive contact binaries in the LMC is 38: 30 in the OGLE+MACHO database, 5 systems in the VFTS-TMBM sample and 3 other miscellaneous ones. Although we do not have the respective contributions of O and B-type stars to the OGLE+MACHO sample, we expect that the B-type contact binaries will dominate the sample owing to their higher lifetime. Assuming thus that less than 50\% of the known LMC contact binaries are O+O-type systems, our estimated number of these systems is in agreement with the data.

These calculations implicitly assume that there are no strong observational biases and that the star formation rate has been constant for a period of time longer than the lifetime of a 15\Msun star. However, \citet{schneider2018a,schneider2018b} suggest that the star formation rate of 30~Dor has been declining over the last 10\,Myr. They also expect a shallower IMF for 30~Dor than the Salpeter one, which may be a reason for the abundance of high mass stars $\geq30$M$_{\odot}$ in the VFTS sample.  These numbers also depend on the power laws we assume for the birth parameters of our synthetic populations.

Extending our calcualtions to the SMC, we first estimate the number of O-type stars in the SMC by comparing its star formation rate of $\sim0.05\Msun$/yr \citep{harris2009} to that of the LMC $\sim0.2\Msun$/yr \citep{hagen2017}. Given thus that the star formation rate of the SMC is 1/4th of the LMC and by assuming f$_\textrm{bin}=0.55$, we expect  a maximum of 2 over-contact or 3 relaxed-contact O-type binaries in the SMC.

\section{Discussion}
\label{discussion}

From our binary evolution models, we find that nuclear-timescale massive contact binaries evolve towards equal masses. This result agrees with the confirmed over-contact systems, which indeed have equal or nearly equal masses ($\textrm{q}\approx1$), thus implying that a mass ratio of unity is the equilibrium configuration that contact binaries tend to achieve. We further infer from our synthetic contact-binary populations that these observed equal-mass systems must have had initial mass ratios $\geq0.8$ and that their fate is to eventually merge on the main sequence. Two of the confirmed over-contact binaries though do not have exactly equal masses--  OGLE SMC SC10-108086 which has $\textrm{q}=0.85\pm+0.06$ and an even more exceptional V729 Cyg which has the lowest mass ratio ($\textrm{q}=0.29$) in our examined sample. The latter though is known to harbour a tertiary companion \citep{kennedy2010, rauw2019}.

Although our predicted mass ratio distribution is heavily skewed towards a value of one, the majority of observed massive contact binaries have unequal masses ($\textrm{q}<0.9$). Upon inspection we find that these unequal mass systems are in fact, either confirmed near-contact binaries or those whose degree of contact is unclear. To study their contribution, we constructed a synthetic population that also include binaries in near-contact, in which both stars have R/R$_\textrm{L}\geq0.9$, i.e., they overflow at least 72\% of their Roche lobe volumes (($\textrm{R}/\textrm{R}_\textrm{L})^3$). While our definition of near-contact binaries is generous, we find that the distributions constructed with these models can accommodate observed systems with mass ratios as low as $\textrm{q}=0.4$. We thus conclude that the observed unequal-mass contact systems are likely to be binaries just nearing contact.

The overall distribution of the 7 LMC spectroscopic contact binaries is better explained by considering the contribution of our near-contact binary models and considering the fact that the VFTS data is skewed towards O-type stars with masses $\geq 15$\Msun.  We make very specific predictions about the distribution of their orbital periods, masses and mass ratios. For over-contact O and early B-type systems in both Magellanic Clouds, i.e., where both stars in the binary have R/R$_\textrm{RL}\geq1$, we expect the total mass of the majority of observed contact binaries to be  $\lessapprox45$\Msun and to have orbital periods less than 1.5\,days, with the most likely period to be between 0.7 and 1\,day.  With a more relaxed definition of contact, i.e., by also including ner-contact binaries, we predict orbital periods to be up to 2\, days. Such low orbital periods are reported for over half of the massive contact systems from the OGLE and MACHO samples \citep{rucinski1999,pawlak2016}. However, as these samples also include late B-type binaries and the individual orbital parameters of the systems are unavailable, it is difficult to isolate their contributions to the sample.

Our predicted distributions indicate a strong correlation between the period and total mass of a contact binary, and also between its period and mass ratio, which we also find in the observed sample of contact binaries-- wider binaries typically have higher masses and lower mass ratios (Figures~\ref{LMC_contact} and \ref{masscut_contact}). Despite this interesting agreement, which is further fortified by including near-contact binaries or by only retaining O-type stars in our synthetic populations, we still find that our predicted mass ratio distribution is far too skewed towards q=1 and, the correlation between the period and total mass of the observed systems (for total masses $\geq40$\Msun, especially in the Galaxy) is not as tight as our predictions. In order to evaluate these findings, let us examine the contribution of the uncertainties in the physics of our binary evolution models and in the observations of massive contact binaries.

As in all massive star models, the main uncertainties relate to internal transport processes and mass loss. However, in contact binaries, some of these processes have a particular emphasis.
One important issue may be that while our employed contact treatment does consider the mechanical
component of a common envelope of both stars within the Roche approximation well, the thermal
component is not dealt with. Despite their common envelope, no energy transport from one star to the other is
considered in our models, and our contact scheme allows both stars to possess different surface temperatures. An argument that this may not cause large errors is that the internal structure of models of stars is largely independent of the outer boundary conditions  \citep{kippenhahn1990}. Besides, the two components of our binary models differ typically by less than 20\% in their effective temperatures, and have a decreasing trend with time as the models tend towards a mass ratio of one. A temperature difference of this order between both components is confirmed by observations of contact binaries (cf., Tables~\ref{MC_contact}, \ref{gal_contact} and, Fig.~\ref{HRD_contact}) and more recently by \citet{abdul-masih2020} for the over-contact binaries, VFTS\,352, V\,382 Cyg and SMC-SC10 108086. They speculate however, that a heat exchange mechanism may be responsible for the near equalization of the surface temperatures of these stars.

Internal chemical mixing is also particularly uncertain in close binaries.  It has been postulated that Eddington-Sweet circulations in tidally-locked binaries can lead to chemically homogeneous evolution (e.g., \citealp{deMink2009, marchant2016, hastings2020}). \cite{hastings2020} showed that the Eddington-Sweet circulations in tidally-locked binaries can be up to twice faster than in a comparable
rotating single star. This effect is not included in our models. However, even with the enhanced Eddington-Sweet circulations, initial total masses above nearly 100\Msun are required for this channel to operate, which is beyond the upper limit
of our model grids. From the observed contact systems, VFTS\,217 is closest to this regime. However, \citet{mahy2020b} find that neither of the components of this system is overluminous, although it should be borne in mind that this system is just approaching contact. 
On the other hand, \citet{abdul-masih2020} show that likely both components, but certainly the less 
massive one, of the Galactic system V382\,Cyg and SMC system OGLE SMC-SC10\,108086 are overluminous. This might argue for
strong internal mixing, however, the surface abundances of the six stars in these three binaries are not strongly enriched \citep{abdul-masih2019, abdul-masih2020}. While we can not draw any firm conclusion here, internal mixing remains only a weak candidate for explaining the discrepancies between our models and some of the observed binaries. 

Mass loss may also be an important process in some of the binary systems considered here. First, in particular while comparing with the most massive and metal-rich contact binaries,  their strong stellar winds must be accounted for. While stellar winds are included in our models, the mass loss rates of Galactic O\,stars are thought to be about 1.6 times higher than of those in the LMC \citep{vink2001, mokiem2007}, which could be relevant for the most massive Galactic contact systems  especially to explain their wide orbits. While we can only speculate how the orbital-parameter distributions of massive Galactic contact systems may behave, a separate grid of MW binary models is required to ascertain the true nature of these distributions. We note that the widest known LMC (near) contact system, VFTS\,217, which is the only clear outlier in the P$_\textrm{contact}$-q$_\textrm{contact}$ diagram of the Magellanic Clouds sample (Fig.~\ref{LMC_contact}), has a total mass above 80\,M$_{\odot}$. Among the MW contact binaries, LY Aur, XZ Cep, V348 Car and V729 Cyg have exceptionally wide periods ($\geq4\,$days) compared to the predicted period domain from our distributions. They have total masses $\gtrapprox40$\Msun, except for XZ Cep which although reported as a contact system, is in fact a semi-detached binary.  

A clue for understanding these wide and massive metal-rich systems  can be found by examining the HRD positions of their components in Fig.~\ref{HRD_contact}. Our models predict that short-period contact systems should be found close to the ZAMS on the HRD, which is in agreement with the majority of the observed systems. This situation is only different for the  Galactic systems with periods $\geq4$\,days, which are much further evolved along the MS. Obviously, here both components keep overfilling their Roche lobes despite the large system dimensions. One reason they may do that could be their proximity to the Eddington limit, which inflates the envelopes of these stars \citep{sanyal2015}. Since envelope inflation is metallicity dependent (\citealp{sanyal2017}, and see Fig.~\ref{HRD_contact}), this effect is most strongly expected in the MW, while in our lower metallicity models, it does not play a significant role.

Mass-transfer related mass loss from binary systems may also be important here. Notably, our model calculations are conservative, in that all the mass transferred from the mass donor is assumed to end up on the mass gainer. It is well known that many mass transferring binaries can
spill a considerable fraction of the transferred matter out of the binary system \citep{demink2007,langer2012}. In fact, one of the Galactic contact binaries, SV Cen does have an accretion disk detected around it \citep{linnel1991, shematovich2017}. While the dependence of this binary mass loss on the system parameters is not well known, and the slow, nuclear timescale mass exchange in contact systems may be close to conservative, the thermal timescale mass transfer which starts the contact binary evolution (cf., Section~\ref{examples})
may have well been non-conservative, especially for the systems with the rather large orbital periods (P$_\textrm{contact}\gtrapprox1.2$\,days) as is the case for some of the most massive observed contact binaries. A possible mode of non-conservative mass transfer in our models may occur via isotropic  re-emission from the mass gainer after it accretes mass \citep{soberman1997}, which may increase the specific angular momentum of the binary and widen its orbit compared to what we will be obtained from a conservative assumption of mass transfer \citep{sen2021, hastings2021}. 

In our study we have not addressed the actual formation of massive binary systems with initial orbital periods within the range considered here, i.e., between 0.6 and 2.0 days. Though beyond the scope of this study, this is a relevant problem as pre-main-sequence (PMS) stars have larger radii relative to ZAMS stars, facilitating interaction during the PMS-phase, and ZAMS component masses are correlated. Studies of lower mass stars consider Kozai-Lidov oscillations and dynamical instability in triple systems as channels toward the formation of close binary systems \citep{sana2017b, moe2018}. Indeed some of the MW contact binaries do have tertiary companions or binary systems detected, which may have also played a role in their current orbital period, inclination and eccentricity. For massive stars, interaction with primordial gas regulated by an external magnetic field \citep{lund2018}, along with disc fragmentation and the subsequent migration of protostars forming in these fragments \citep{meyer2018} have been proposed as well. That orbital hardening in general is an essential ingredient of close binary formation is supported by recent findings by \citet{ramirez2021}, who find evidence for migration by comparing binary properties in young OB associations. Finally, we stress that our assumptions for the initial distribution of periods and mass ratios are subject to uncertainties. Observations of systems with masses and periods of interest are scarce. They are typically derived from OB associations and clusters that are estimated to be 0--4 Myrs old. Some systems may have already undergone some evolution and mass exchange by the time we can observe them (such as binaries that evolve like our example system 1). Our predictions for the observable P-q distribution are therefore likely somewhat sensitive to the precise choices made.

Overall, we can say that our contact binary evolution models appear to reproduce the observed properties of contact binaries in the lower total mass range (below about 45\Msun) reasonably well, particularly for the metallicities of the Magellanic Clouds. Our models however,  appear to overestimate the fraction of contact binaries with mass ratios close to one and fall short of explaining the scatter in the period--total mass distribution of the more massive contact binaries, especially in the MW.

\section{Conclusions}
\label{conclusions}

We have performed the first dedicated investigation of the evolution of massive contact binaries using our current theoretical understanding of this phase and made predictions for their observed population. We have computed two large grids of detailed binary evolution models for the LMC and SMC metallicities in \texttt{MESA}, by assuming mass transfer to be conservative and by using a contact prescription that considers the mechanical consequences of a common envelope within the Roche model, but neglects any energy exchange between the components of the contact binary. 

From our models, we find that binaries that are in contact over nuclear timescales will inevitably merge. Hence we predict that all the observed contact binaries will eventually merge on the main sequence. From our models we find that a stellar merger is the outcome for most binaries in the Magellanic Clouds with initial periods less than 2\,days and a definitive fate for all binaries with periods less than 1.5\,days. This contact duration can last up to 10.7\,Myrs in the lowest mass binaries in our grid, before they ultimately merge.

Our nuclear-timescale contact binary models evolve towards equal component masses, corresponding to the lowest energy state of binary systems \citep{adams2020}. The longer a system spends time in contact, the likelier it attains a mass ratio of 1 on the main sequence. 
Nearly all the observed systems in deep contact and some in near contact have mass ratios between 0.9 and 1, and hence agree well with our result. From our models, we infer that these contact binaries would have had initial mass ratios $\geq0.8$. We also expect massive binaries to spend less time in contact as their initial mass or initial period increases, and when their initial mass ratio deviates further from 1. None of our contact binary models undergo chemically homogeneous evolution in both grids.

We expect nearly 4.5\% of the O-star binaries in the LMC and SMC to be over-contact binaries (10 systems in total), assuming a binary fraction of $\approx 50\%$ and constant star formation. With a binary fraction of 100\%, the number of over-contact binaries will be 1.6 times as many and even more if we include binaries in near-contact (R/R$_\textrm{RL}\geq0.9$ for both stars).

We find it essential to consider these near-contact binaries in our study due to the observational uncertainties in the degree of contact and in the configuration of reported contact binaries. We also find that their inclusion while computing the distribution of orbital period and mass ratio, reproduces the observed distribution more closely. In a population of O \& early B-type binaries in the Magellanic Clouds, we expect that the contact binaries will have total mass $\lessapprox45\Msun$, periods  $<1.5$\,days and mass ratios $\geq0.9$. By including near-contact binaries, our distributions predict a wider domain of probable orbital periods of up to 2\,days and of mass ratios as low as 0.4. From our modelled distributions, we predict that all reported contact binaries which have mass ratios $\lessapprox0.80$ and periods greater than 1\,day, are likely to be near-contact systems.


 We find little difference in the stellar and orbital-parameter distributions between our LMC and SMC contact binary populations, implying that metallicity does not play an important role in their orbital evolution at the values we have considered. Significant differences only appear in the orbital period distribution if we place mass cuts on our synthetic populations, by excluding systems with total masses less than 30\Msun. 

Our binary models identify a distinct subspace in the orbital period-mass ratio diagram 
in which most contact systems are expected (Fig.~\ref{LMC_contact}), which we find indeed populated
by most of the observed Magellanic Cloud contact binaries with total masses below 80\Msun. Like the general trend seen in the observed data, our models also predict that wider binaries tend to be more massive and have small mass ratios. Overall, this argues for our treatment of the contact phase to capture the essential mechanics of the orbital evolution of the binaries during this phase, especially for systems with total masses $\geq45$\Msun. For higher mass systems we cannot
reproduce the individual values of their mass ratios or orbital periods, especially when we consider the Galactic systems. This is mainly due to the fact that  lower-mass systems are weighted higher in a distribution as they are favoured more by the initial mass function and their overall longer lifetimes. While imposing a mass cut on the distribution or including near-contact binaries does somewhat alleviate the problem, other physics phenomena may also be at play in the evolution of these massive systems such as stellar winds, non-conservative mass transfer and envelope inflation, which may be especially relevant for the Milky Way binaries.



To test these ideas, we need to compute dedicated model grids for the Milky Way to compare with the Galactic massive contact binaries. In Paper II, we intend to explore these avenues and compare the stellar parameters of our models with the currently observed massive contact binaries, such as their luminosities, effective temperatures, rotational velocities and surface abundances.

We conclude thus that the contact phase is an inevitable, nuclear-timescale evolutionary phase for most massive binary stars born with periods less than 2\,days. We also identify it an observable phase preceding an imminent stellar merger, lending us the opportunity to probe the conditions that lead to one of the least understood evolutionary phases in stellar physics. While stellar mergers themselves produce exotic and peculiar stellar phenomena and supernovae, the contact phase in extremely metal-poor and massive binaries has also been flagged as a channel to produce the progenitors of merging black holes \citep{marchant2016}. Therefore we hope that, ultimately, our results may not only lead to a better understanding of massive contact binaries in general, but also lead to better predictions of progenitors of stellar mergers and of massive compact object mergers, not only of those in our cosmic neighbourhood but also those which live so far out in the Universe that direct observations may remain difficult for the foreseeable time.

\section*{Acknowledgements}
This project was funded in part by the European Union’s Horizon 2020 research and innovation program from the European Research Council (ERC, Grant agreement No. 715063,  772225: MULTIPLE), and by the Netherlands Organization for Scientific Research (NWO) as part of the Vidi research program BinWaves with project number 639.042.728 and the National Science Foundation under Grant No. (NSF grant number 2009131). D.Sz. is supported by the Alexander von Humboldt Foundation. PM acknowledges support from the FWO junior postdoctoral fellowship No. 12ZY520N.
SdM is grateful to all members of the BinCosmos group for enlightening discussions and input, in particular to Lieke van Son and Katie Sharpe.







\appendix

\section{Additional material}
\subsection{Probability distributions of contact systems according to their initial parameters}
\label{initial_dist}

In this section we elaborate how we compute the distribution of contact systems as a function of their initial parameters: $\textrm{P}_\textrm{i}, \textrm{q}_\textrm{i}$ and $\textrm{M}_\textrm{T,i}$. To compute how contact systems are distributed as a function of $\textrm{P}_\textrm{i}$, we first compute the weighted contact duration of each system and then the contribution of that system towards a given $\textrm{P}_\textrm{i}$: 

\begin{align}
\textrm{H}_\textrm{contact}(\textrm{P}_\textrm{i})=\delta_{\textrm{P}_\textrm{i}} \frac {w_\textrm{s}\tau_\textrm{contact,s}} {\sum_{\textrm{s=1}}^{\textrm{n}} w_\textrm{s}\tau_\textrm{contact,s}} \label{eq:X_Pi} \end{align} 

where, P$_\textrm{i} \,\epsilon\, [0.6, 0.7, 0.8, ...,2.0]$ days and $\delta_{\textrm{P}_\textrm{i}}$ takes the value of either 0 or 1. For e.g., for systems with an initial period of 1\,day,  $\delta_{\textrm{P}_\textrm{1.0}}$; for all other systems it takes the value of 0.  $\tau_\textrm{contact,s}$ is the net time spent in contact by the system and w$_\textrm{s}$ is the birth weight of the system as calculated in Eqs.~\ref{eq:weight}. 

We compute similar functions for each q$_\textrm{i}$ and M$_\textrm{T,i}$ in our grid. These are later used to make 2D probability distributions 
of  P$_\textrm{i}$ vs. q$_\textrm{i}$ and P$_\textrm{i}$ vs. M$_\textrm{T,i}$ (as will be seen in Fig.~\ref{LMC_contact}).

\subsection{Probability distributions of contact systems according to observed parameters}
\label{contact_dist}

In this section we show how we compute the distribution of contact systems according to their present-day parameters.

We divide the whole range of orbital periods for contact systems during their evolution in bins of 0.1 days. We take each system from the grid and determine how much time it spends in contact in each bin and weigh this time by its initial weight. To construct the distribution of the observed orbital period, we sum this value in each period bin and normalize its contribution to the total weighted contact duration of the entire grid ($\textrm{H}_\textrm{contact}(\Delta \textrm{P})$). 

\begin{align}
\textrm{H}_\textrm{contact}(\Delta \textrm{P}) = \delta_{\Delta \textrm{P}} \frac {\sum_{\textrm{s=1}}^{\textrm{n}}  w_\textrm{s}\,\textrm{dt}_\textrm{s,$\Delta$P}}{\sum_{\textrm{s=1}}^{\textrm{n}}w_\textrm{s}\,\textrm{dt}_\textrm{s,$\Delta$P}} \label{eq:X_contact}
\end{align} 

where, w$_\textrm{s}$ is the birth weight of the system,  $\Delta$P is the orbital period bin of systems while they are in contact such that $\Delta$\,P $\,\epsilon\, [0.6-0.7, 0.7-0.8, ...,3.0-3.1]$ days, $\textrm{dt}_\textrm{s,$\Delta$ P}$ is the time spent by the system while its in contact in each $\Delta$\,P bin. $\delta_{\Delta \textrm{P}}$ is either 0 or 1; if a system spends time in contact in the bin $\Delta \textrm{P}=0.6-0.7$\,days, then $\delta_{0.6-0.7}=1$ else it is 0. 

From the distribution thus constructed, we can compute the probability of finding a contact system with a particular current orbital period and hence their most likely observed period.  We similarly compute H$_\textrm{contact,$\Delta$q}$ and 
H$_\textrm{contact,$\Delta$M$_\textrm{T}$}$ and the 2D probabilities of finding contact systems for combinations of mass ratio and total mass.

\begin{figure*}
   \centering
   \includegraphics[angle=90, width=0.7\linewidth, clip, ]{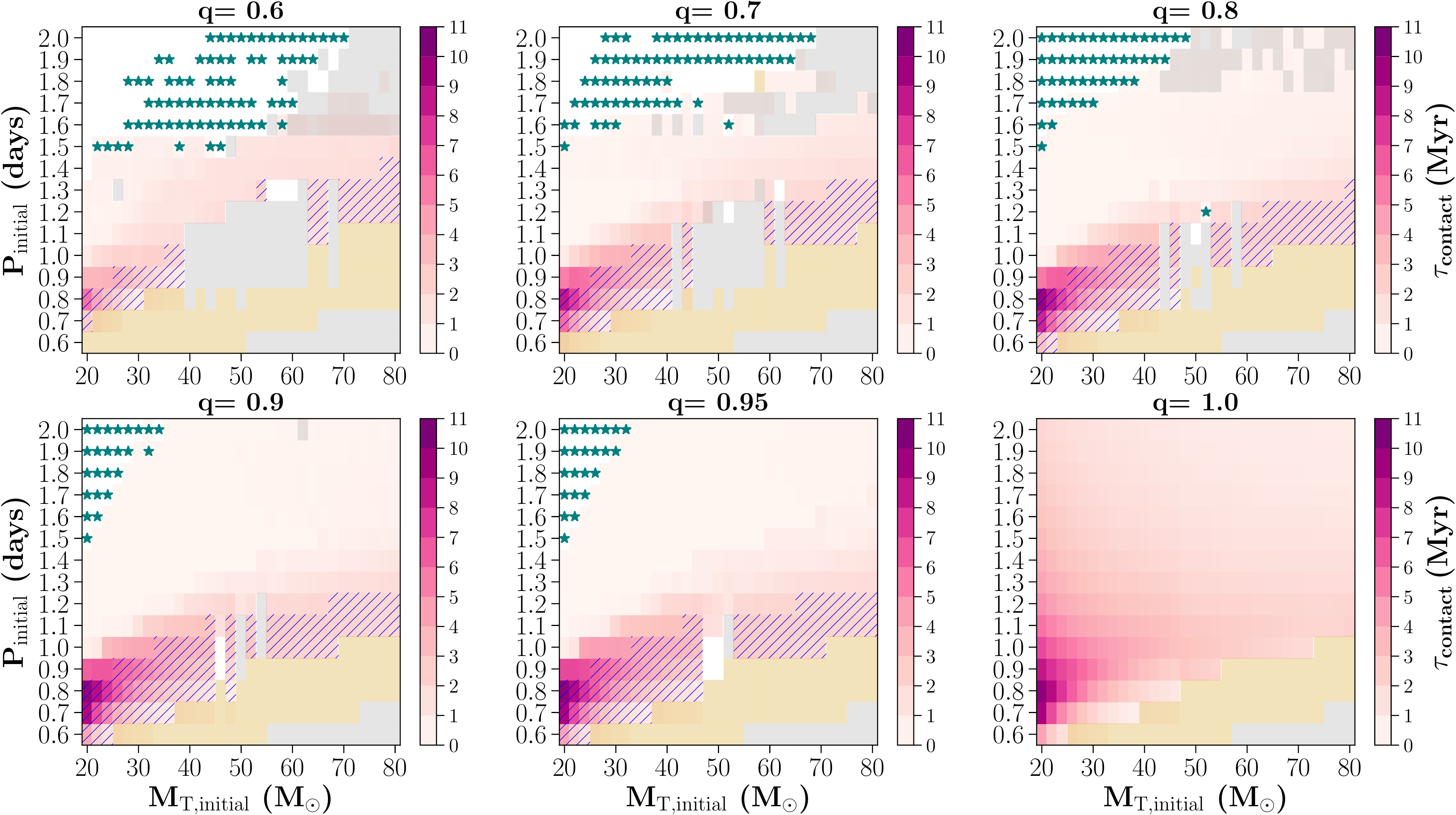}
    \caption{The contact duration ($\tau_\mathrm{contact}$) of each system  of the entire LMC grid in a P$_\mathrm{i}$--M$_\mathrm{T,i}$ for all q$_\mathrm{i}$. Background colors and symbols are as explained in Fig.~\ref{scatter_sub}.}
    \label{scatter_contact_LMC}
\end{figure*}

\begin{figure*}
\centering
\begin{subfigure}{0.5\linewidth}
\centering
  \includegraphics[width=\linewidth,left]{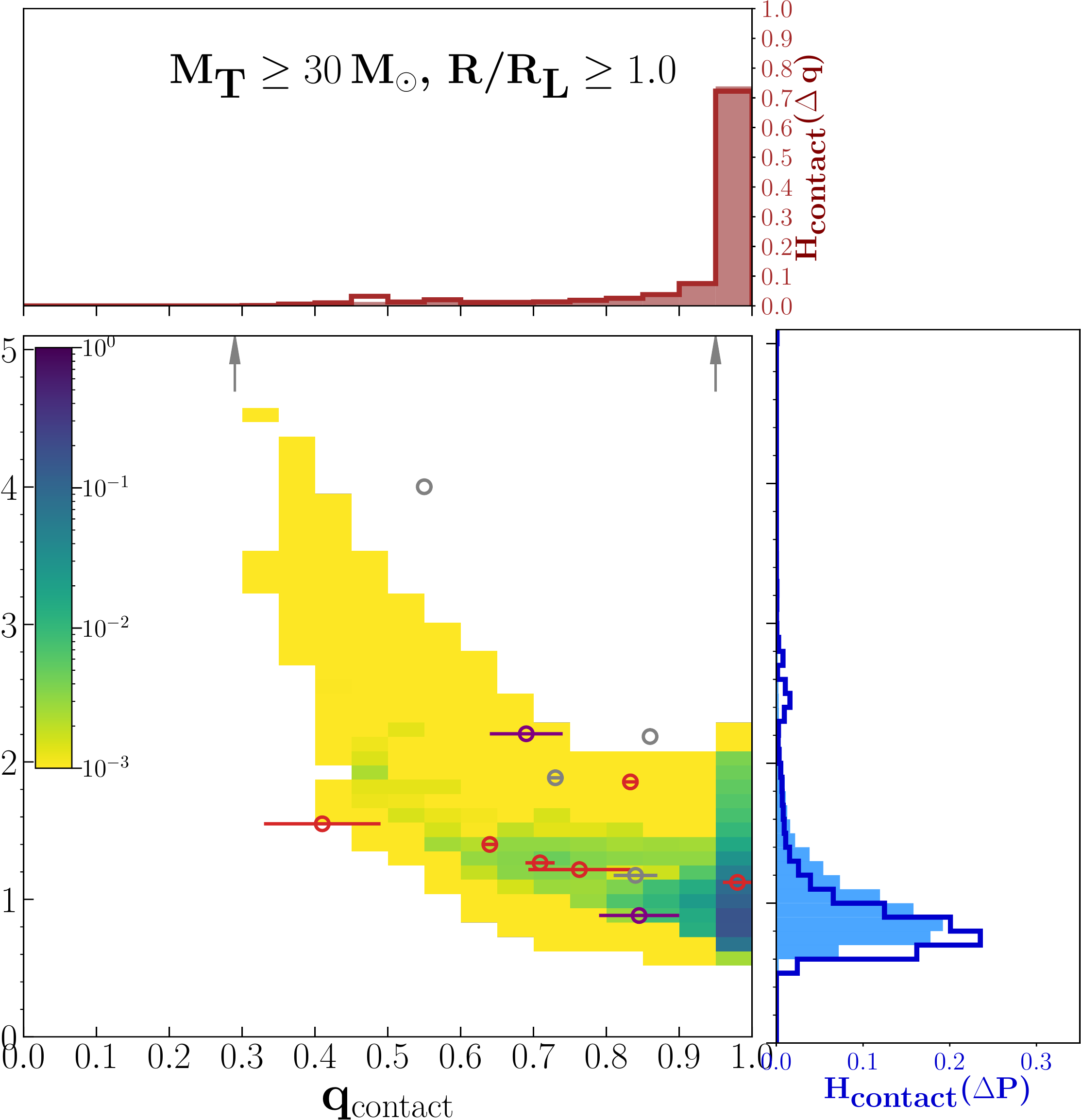}
\end{subfigure}%
\begin{subfigure}{0.5\linewidth}
  \centering
  \includegraphics[width=\linewidth,clip]{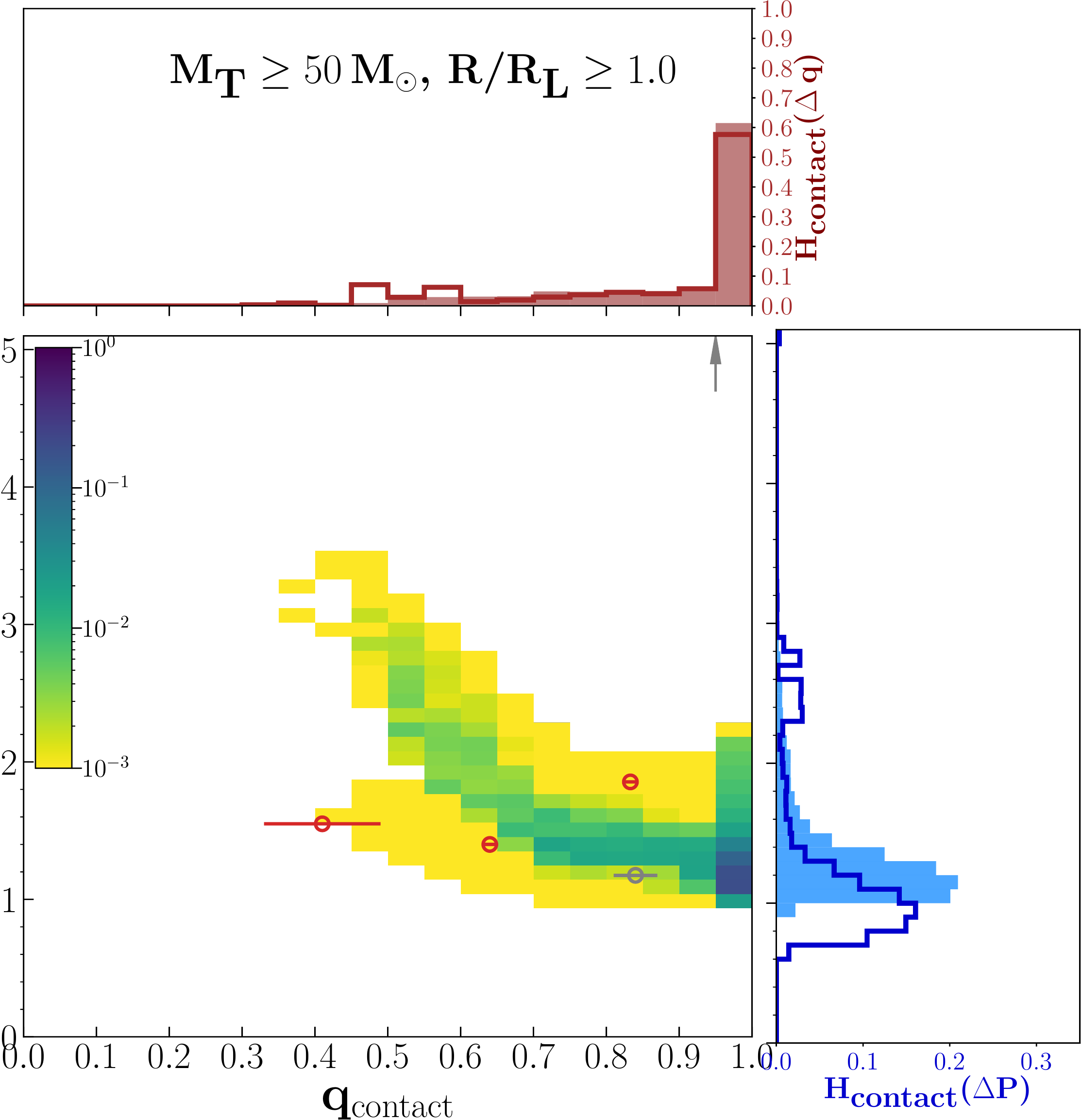}
\end{subfigure}
\caption{The 2D distribution of P$_\textrm{contact}$ vs. q$_\textrm{contact}$ for contact and over-contact binaries, with mass cuts on initial total mass: M$_\textrm{T,i}\geq30\Msun$ (left) and M$_\textrm{T,i}\geq50\Msun$ (right). Colors, histograms and markers are as explained in Fig.~\ref{LMC_contact}.}
\label{masscut_contact}
\end{figure*}

\section*{Data availability}
The template folders to run our single star and binary models used in this work are available at :  \url{https://github.com/athira11/massive_contact_binaries.git}. If they are used, please cite this work as the source of the data. If particular models are required, please send an email to the first author.

\clearpage
\bibliographystyle{mnras}
\bibliography{master} 
\bsp	
\label{lastpage}
\end{document}